\documentclass[twocolumn]{aastex62}

\hypersetup{linkcolor=red,citecolor=blue,filecolor=cyan,urlcolor=magenta}

\newcommand{\kms}{\mbox{km\,s$^{-1}$}}
\newcommand{\methanol}{\mbox{$\rm CH_3OH$}}
\newcommand{\ntwodp}{\mbox{$\rm N_2D^+$}}
\newcommand{\ntwohp}{\mbox{$\rm N_2H^+$}}
\newcommand{\ceighteeno}{\mbox{$\rm C^{18}O$}}
\newcommand{\thirteenco}{\mbox{$\rm ^{13}CO$}}
\newcommand{\dcop}{\mbox{$\rm DCO^+$}}
\newcommand{\hcop}{\mbox{$\rm HCO^+$}}
\newcommand{\dcn}{\mbox{$\rm DCN$}}
\newcommand{\ammonia}{\mbox{$\rm NH_3$}}
\newcommand{\gcm}{\mbox{g cm$^{-2}$}} 
\newcommand{\mjypbm}{\mbox{mJy\,beam$^{-1}$}}
\newcommand{\jypbm}{\mbox{Jy\,beam$^{-1}$}}

\received{}
\revised{}
\accepted{}

\shorttitle{Gas Kinematics of a Massive Protocluster}
\shortauthors{Cheng et al.}

\begin{document}

\title{GAS KINEMATICS OF THE MASSIVE PROTOCLUSTER G286.21+0.17 REVEALED BY ALMA}

\author{Yu Cheng}
\affil{Dept. of Astronomy, University of Virginia, Charlottesville, Virginia 22904, USA}

\author{Jonathan C. Tan}
\affil{Dept. of Space, Earth \& Environment, Chalmers University of Technology, Gothenburg, Sweden}
\affil{Dept. of Astronomy, University of Virginia, Charlottesville, Virginia 22904, USA}

\author{Mengyao Liu}
\affil{Dept. of Astronomy, University of Virginia, Charlottesville, Virginia 22904, USA}

\author{Wanggi Lim}
\affil{SOFIA-USRA, NASA Ames Research Center, MS 232-12, Moffett Field, CA 94035, USA}

\author{Morten Andersen}
\affil{Gemini Observatory, NSFs National Optical-Infrared Astronomy Research Laboratory Casilla 603, La Serena, Chile}



\begin{abstract}
We study the gas kinematics and dynamics of the massive protocluster
G286.21+0.17 with the Atacama Large Millimeter/submillimeter Array
using spectral lines of \ceighteeno(2-1), \ntwodp(3-2), \dcop(3-2) and \dcn(3-2).
On the parsec clump scale, \ceighteeno{} emission appears highly filamentary around 
the systemic velocity. 
 \ntwodp{} and  \dcop{}  are more closely associated with the dust continuum.
 \dcn{}  is  strongly  concentrated  towards  the protocluster center, 
 where no or only weak detection is seen for \ntwodp{} and \dcop, 
 possibly due to this region being at a relatively evolved evolutionary stage.
Spectra of 76 continuum defined dense cores, typically a few 1000 AU in size, are analysed to measure
their centroid velocities and internal velocity dispersions. There are 
no statistically significant velocity offsets of the cores among the different dense gas tracers. Furthermore, the majority (71\%) of the dense cores have subthermal velocity offsets 
with respect to their surrounding, lower density \ceighteeno{} emitting gas.
Within the uncertainties, the dense cores in G286 show internal kinematics that are consistent with being in virial equilibrium. 
On clumps scales, the core to core velocity dispersion is
also similar to that required for virial equilibrium in the protocluster
potential. However, the distribution in velocity of the cores is largely composed of two
spatially distinct groups, which indicates that the dense
molecular gas has not yet relaxed to virial equilibrium, perhaps due to there being recent/continuous infall into the system.
\end{abstract}


\section{Introduction}\label{sec:intro}

While it is generally agreed that most stars form in clusters and/or
associations rather than in isolation
\citep[e.g.,][]{Lada03,Gutermuth09,Bressert10}, there is no consensus
for how this comes about.
Several fundamental questions about star cluster formation are still
debated. For example, is the process initiated by internal processes
within a Giant Molecular Cloud (GMC), such as decay of support by
supersonic turbulence or magnetic fields, or external processes, such
as triggering by cloud-cloud collisions or feedback-induced shock
compression \citep[see e.g.,][]{Tan15}.

Once underway, is cluster formation a fast or a slow process relative
to the local freefall time ($t_{\rm ff}$)?  \citet{Tan06} and
\citet{Nakamura07} proposed that formation times are relatively long,
i.e., $\sim 10 t_{\rm ff}$, especially for those clusters with high
($\gtrsim$ 30\%) overall star formation efficiency, since simulations
of self-gravitating, turbulent, magnetized gas show low formation
efficiency of just $\sim$2\% per free-fall time
\citep{Krumholz05,Padoan11}.  Alternatively, \citet{Elmegreen07},
\citet{Hartmann07} and \citet{Hartmann12} have argued for cluster
formation in just one or a few free-fall times. Another question is:
what sets the overall star formation efficiency during cluster
formation? The formation timescale and overall efficiency are likely
to affect the ability of a cluster to remain gravitationally bound,
which on large scales influences global ISM feedback, e.g.,
concentrated feedback from clusters can create superbubbles
\citep[e.g.,][]{Krause13}, and on small scales controls the feedback
environments and tidal perturbations of protoplanetary disks
\citep[e.g.,][]{Adams10}.


Star cluster formation is likely to be the result of a complex
interaction of numerous physical processes including turbulence,
magnetic fields and feedback.  From the observational side, measuring
the structure and kinematic properties of the dense gas component is
needed to provide constraints for different theretical
models. Previously, \citet{Walsh04} found small velocity differences
between dense cores and surrounding envelopes for a sample of low-mass
cores. \citet{Kirk07,Kirk10} surveyed the kinematics of over 150
candidate dense cores in the Perseus molecular cloud with pointed
\ntwohp{} and \ceighteeno{} observations and found subvirial core to
core velocity dispersions in each region.
A similar small core velocity dispersion was also found in the
Ophiuchus cloud \citep{Andre07}.  \citet{Qian12} searched for
\thirteenco{} cores in the Taurus molecular cloud and found the core
velocity dispersion exhibits a power-law behavior as a function of the
apparent separation, similar to Larson’s law for the velocity
dispersion of the gas, which suggests the formation of these cores has
been influenced by large-scale turbulence.

These observations have generally focused on nearby low-mass
star-forming regions.  With the unprecedented sensitivity and spatial
resolution of ALMA, more light has been shed on massive star forming
regions from the ``clump'' scale (of about a few parsecs) to the
``core'' scale ($\sim0.01$ to $0.1\:$pc)
\citep[e.g.,][]{Beuther17,Fontani18,Lu18}. Multiple coherent velocity
components from filamentary structures have been reported in some
massive Infrared Dark Clouds (IRDCs)
\citep{Henshaw13,Henshaw14,Sokolov18}, similar to the structures seen
in the nearby Taurus region by \citet{Hacar13}. ``Hub-filament''
systems have also been reported in some massive star forming regions
across a variety of evolutionary stages, perhaps indicating presence
of converging flows that channel gas to the junctions where star
formation is most active
\citep[e.g.,][]{Hennemann12,Peretto14,Lu18,Yuan18}.

However, complete surverys for the dense gas component of massive
protoclusters down to the individual core scale, are still rare
\citep[e.g.,][]{Ohashi16,Ginsburg17} and a large spatial dynamic range
is required to perform a multi-scale kinematics analysis.


Until recently only very few nearby regions were known that were
candidates for very young and still forming massive star clusters.
One particular promising star-forming clump is G286.21+0.17 (in short
G286).  It is a massive protocluster associated with the $\eta$ Car
giant molecular cloud at a distance of 2.5 $\pm$ 0.3 kpc, in the
Carina spiral arm \citep[e.g.,][]{Barnes10}.  We performed a core mass
function (CMF) study towards this region based on ALMA Cycle 3
observations in \citet{Cheng18}.

Here we present a follow-up study of multiple spectral lines to
investigate the gas kinematics and dynamics of G286 from clump to core
scales.  The paper is organized as follows: in \autoref{sec:obs} we
describe the observational setup and analysis methods; the results are
presented in \autoref{sec:res}. We discuss the kinematics and dynamics for parsec-scale filaments and dense cores separately in \autoref{sec:fila} and \autoref{sec:kine}, and then summarize our findings in \autoref{sec:conclusion}.

\section{Observational Data}\label{sec:obs}
\subsection{ALMA Observations}

\begin{figure*}[ht!]
\epsscale{1.0}\plotone{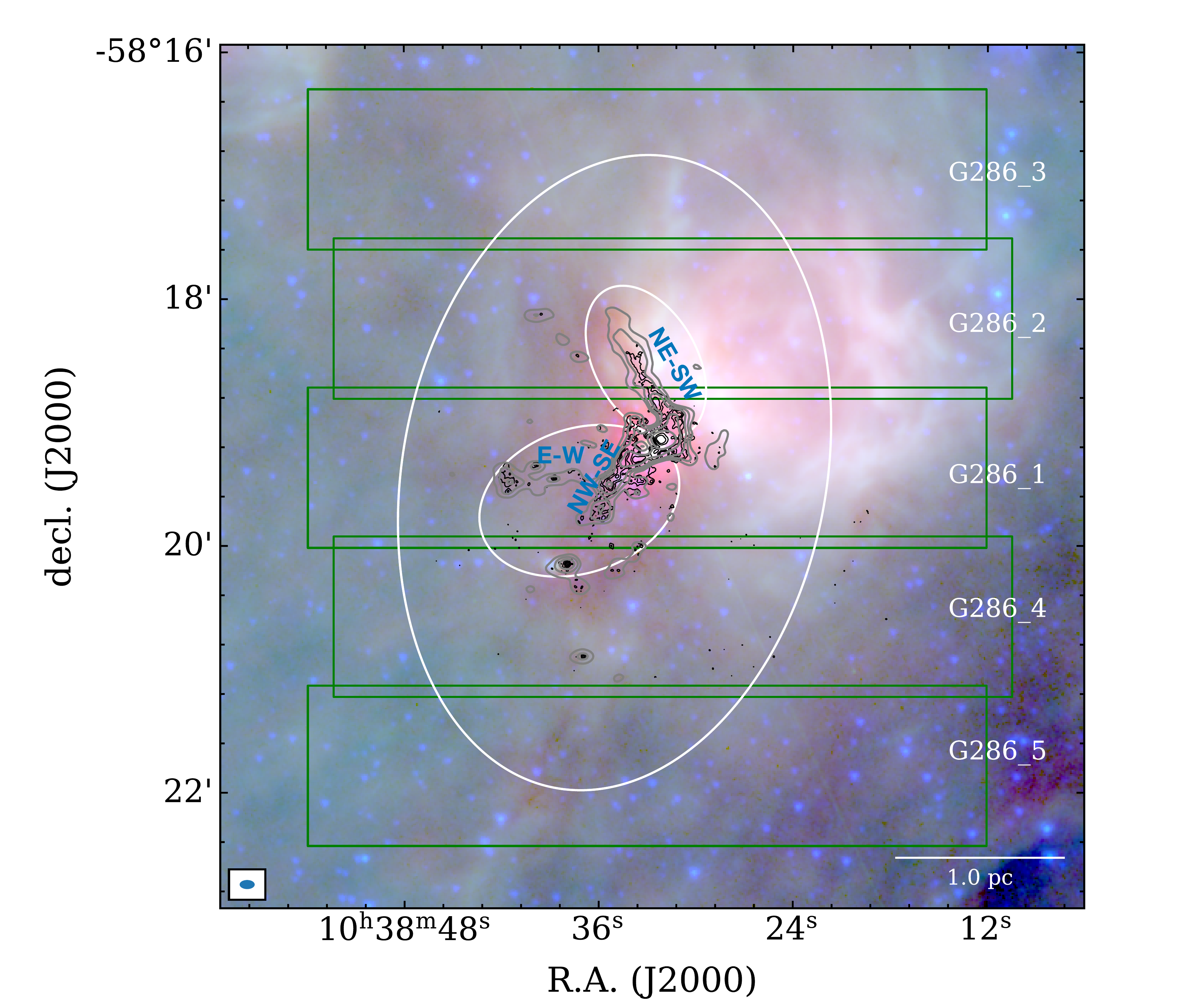}
\caption{
Three color image of G286 constructed by combining Spitzer IRAC 3.6
$\mu$m (blue), 8.0 $\mu$m (green), and Herschel PACS 70 $\mu$m
(red). Black contours show the 1.3 mm continuum image combining ALMA 12-m and 7-m array data (with a resolution of 1.62\arcsec$\times$1.41\arcsec). The contour levels are 1$\sigma$ $\times$ (4, 10, 20, 50, 100) with $\sigma$=0.45\mjypbm. Grey contours show the 1.3 mm continuum image with only 7-m array data (with a resolution of 7.32\arcsec$\times$4.42\arcsec, shown in lower left corner). The contour levels are 1$\sigma \times$ (4, 10, 20, 50, 100) with $\sigma$=1.7\mjypbm. The position of three filamentary structures are marked in blue text. 
The G286 field is divided into five strips,
as shown by the green rectangles. Each strip is covered with 147
pointings of the 12-m array. The large white ellipse denotes the boundary
defined by Mopra HCO$^+$(1-0) emission \citep{Barnes11}, with the
major and minor axes equal to twice the FWHM lengths of the 2D gaussian fits to its emission. Two smaller ellipses indicate the approximate boundaries of two sub-regions with distinct radial velocities revealed by multiple gas tracers.
}
\label{fig:overview}
\end{figure*} 

The observations were conducted with ALMA in Cycle 3 (Project ID
2015.1.00357.S, PI: J. C. Tan), during a period from Dec. 2015 to
Sept. 2016. More details of the observations can be found in
\citet{Cheng18}. In summary, we divided the region into five strips,
denoted as G286\_1, G286\_2, G286\_3, G286\_4 and G286\_5, each about
$1\arcmin\:$ wide and $5.3\arcmin\:$ long and containing 147 pointings
of the 12-m array (see \autoref{fig:overview}).  The position of field
center is R.A.=10:38:33, decl.=-58:19:22.  We employed the compact
configuration C36-1 to recover scales between 1.5\arcsec\ and
11.0\arcsec.  This is complemented by observations with the ACA array,
which probes scales up to 18.6\arcsec.  Total power (TP) observations were
also carried out to recover the total flux (of line emission), which
gives a resolution of about 30\arcsec.

During the observations, we set the central frequency of the
correlator sidebands to be the rest frequency of the
$\rm{N_2D}^+$(3-2) line at $231.32\:$GHz for SPW0, and the
$\rm{C^{18}O}$(2-1) line at $219.56\:$GHz for SPW2, with a velocity
resolution of 0.046 and 0.048~\kms, respectively. The second baseband
SPW1 was set to $231.00\:$GHz, i.e., 1.30~mm, to observe the continuum
with a total bandwidth of $2.0\:$GHz, which also covers CO(2-1) with a
velocity resolution of 0.64~\kms. The frequency coverage for SPW3
ranges from 215.85 to $217.54\:$GHz to observe DCN(3-2), DCO$^+$(3-2),
SiO($v=0$)(5-4) and $\rm{CH_3OH}(5_{1,4}-4_{2,2})$.  This paper will
focus mostly on dense gas tracers \ceighteeno{}, \ntwodp, \dcop{}
and \dcn{}.
 
The raw data were calibrated with the data reduction pipeline using
{\it{Casa}} 4.7.0. The continuum visibility data were constructed with
all line-free channels. We performed imaging with {\it tclean} task in
{\it Casa} and during cleaning we combined data for all five strips to
generate a final mosaic map.  Two sets of images were produced for
different aspects of the analysis, one including the TP and 7-m array
data and one combining TP, 7-m and 12-m data.  The 7-m array data was
imaged using a Briggs weighting scheme with a robust parameter of 0.5,
which yields a resolution of $7.32\arcsec\times4.42\arcsec\:$. For the
combined data, we used the same Briggs parameter. In addition, since
we have extra $uv$ coverage for part of the data, we also apply a
0.6\arcsec~ {\it uvtaper} to suppress longer baselines, which results
in $1.62\arcsec\times1.41\arcsec\:$ resolution.  Both image sets are
then feathered with the total power image to correct for the missing
large scale structures.  Our sensitivity level is about 30 mJy per
beam per 0.1 km/s for $\rm{N_2D^+}$ and $\rm{C^{18}O}$.  A sensitivity
of 45 mJy per beam per 0.1 km/s is achieved for \dcop(3-2), \dcn(3-2),
SiO(5-4) and \methanol$(5_{1,4}-4_{2,2})$.

\subsection{Herschel Observations}


The FIR dust continuum images of G286 were taken from
{\it Herschel} Infrared GALactic plane survey \citep[Hi-GAL;][]{Molinari10,Molinari16}.
The data includes Photodetector Array Camera and 
Spectrometer (PACS) (70 and 160 $\mu$m) and Spectral and Photometric Imaging REceiver (SPIRE) (250, 350, and 500 $\mu$m) images.
We performed pixel by pixel graybody fits to derive the mass surface 
density ($\Sigma$) of the G286 region, following the procedures in \citet{Lim16}.
The background was estimated as the median intensity value between 2 and 4 
times the ellipse aperture shown in \autoref{fig:overview}. To better probe the smaller, higher $\Sigma$ structures, we generated a higher-resolution $\Sigma$ map by regridding the $\lambda\sim$160 to 500$\mu$m images to match the 250$\mu$m data \citep[see][for details]{Lim16}.

 
\section{General Results}\label{sec:res}

An overview of the observed region and the layout of the ALMA observations
is shown in \autoref{fig:overview}.
With the large spatial dynamic range of the ALMA dataset, we will
present the large scale structures traced with single dish TP
observations first, followed by higher resolution 7-m and 12-m array
observations.

\subsection{Observations with the Total Power (TP) Array}

\begin{figure}[ht!]
\epsscale{1.}\plotone{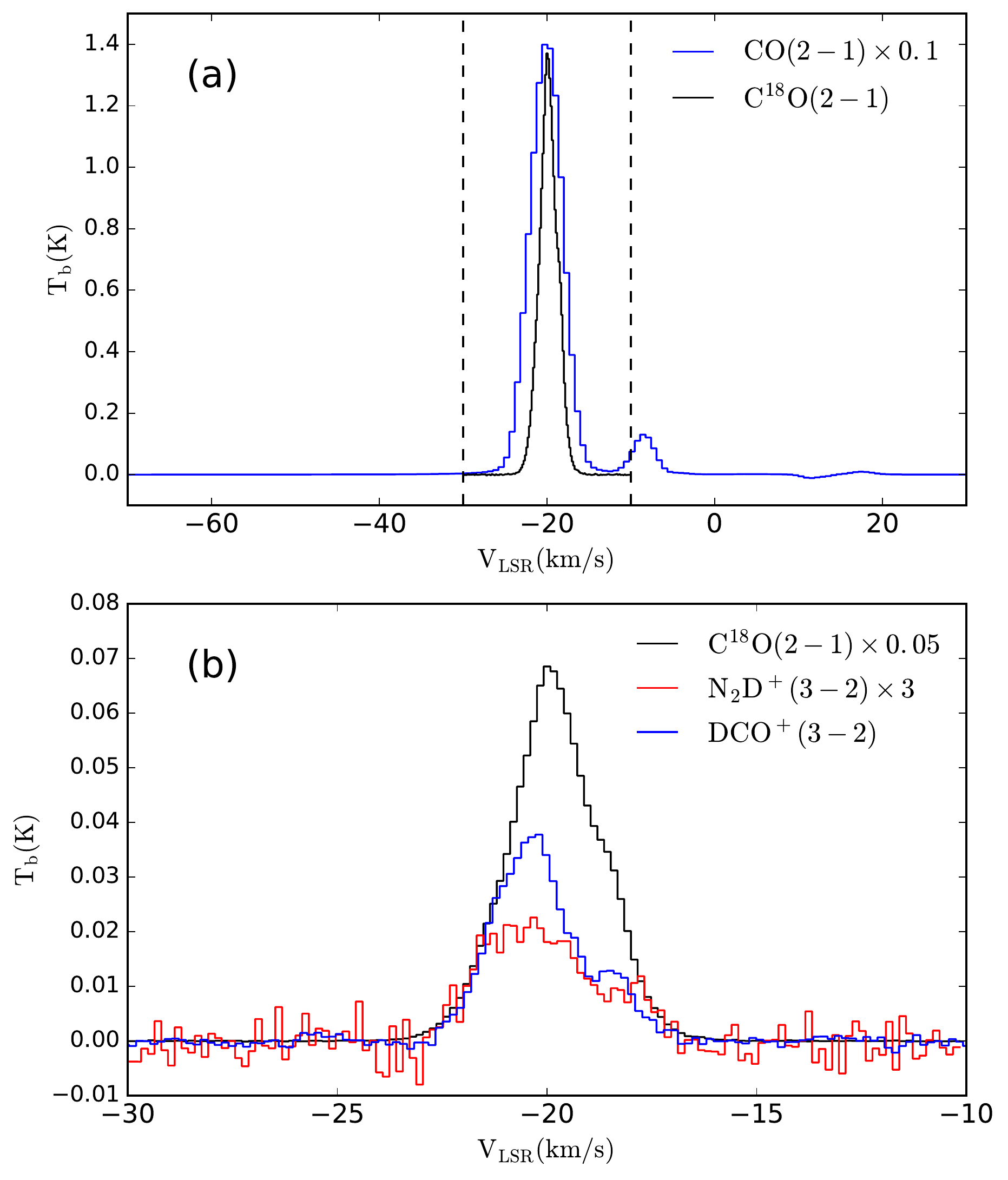}
\caption{
{\it (a)} Averaged CO(2-1) and C$^{\rm 18}$O(2-1) TP spectra extracted
over a 2.5\arcmin~ radius aperture centered on the phase center.  Note
the flux scale of CO(2-1) has been reduced by a factor of 10. {\it
  (b)} Same as {\it (a)} but for \ceighteeno{}(2-1), \ntwodp{}(3-2)
and \dcop{}(3-2) in a smaller velocity range from -30 to -10 \kms.
The flux scale of \ceighteeno{}(2-1) is reduced by a factor of 20 and
that of \ntwodp{}(3-2) is increased by a factor of 3 for ease of
comparison. Note the \ntwodp{}(3-2) emission is affected by hyperfine
structure, while the \dcop{}(3-2) is not.}
\label{fig:tp_spec}
\end{figure}

\autoref{fig:tp_spec}a shows the spectra of CO(2-1) and
\ceighteeno{}(2-1) averaged inside a 2.5\arcmin\ radius aperture
centered on the phase center.  The CO(2-1) line has a maximum around
the known systemic velocity of about -20~\kms.  A secondary, much
weaker peak is seen around -9~\kms. This component is also seen in the
Mopra CO(2-1) map, which appears to be a diffuse structure larger than
our field of view. We expect that this feature is probably contributed
by a foreground or background cloud along line of sight and there is
no indication of an interaction between this cloud and G286. Emission
from $\rm C^{18}O$(2-1) is only seen from the main $-20$~\kms\
component.

\autoref{fig:tp_spec}b shows the spectra of the deuterated dense gas
tracer
\ntwodp{}(3-2) and
\dcop{}(3-2),
averaged over the same region, and compared to \ceighteeno{}(2-1),
zooming-in to the velocity range of the main -20~\kms\ component.
Deuterated species, such as \dcop{} and \ntwodp{}
are expected to be tracers of cold, dense gas, including material that
is contained in pre-stellar cores
\citep[e.g.,][]{Crapsi05,Bergin07,Kong15}, and typically optical thin
even at the core scale, as found in some examples in IRDCs
\citep{Tan13}.
Interestingly, the \ceighteeno{}(2-1) line exhibits a main gaussian-like
profile with a slight skewness (or second component) to the redshifted
side.  The spectrum from \dcop{}(3-2) shows a more pronounced
double-peaked profile, with one component at about -20.5~\kms\ and the
other at -18.5~\kms. 

The double-peak profile, i.e., with a stronger blue wing, has also
been seen in the \hcop{}(1-0) and \hcop{}(4-3) line in
\citet{Barnes10}, with similar central velocities for both peaks. It
was interpreted by \citet{Barnes10} as a canonical inverse P-Cygni
profile indicating gravitational infall \citep{Zhou93}. However, in
this picture we would expect a single gaussian profile for optical
thin tracers at the self-absorption velocity, in contrast to our
\dcop{}(3-2) spectrum. We will return in \autoref{sec:kine} to the
question of whether the claimed inverse P-Cygni profile in \hcop{} is
really tracing global clump infall or whether it is arising from
distinct spatial and kinematic substructures in the protocluster.

\begin{figure*}[ht!]
\epsscale{1.0}\plotone{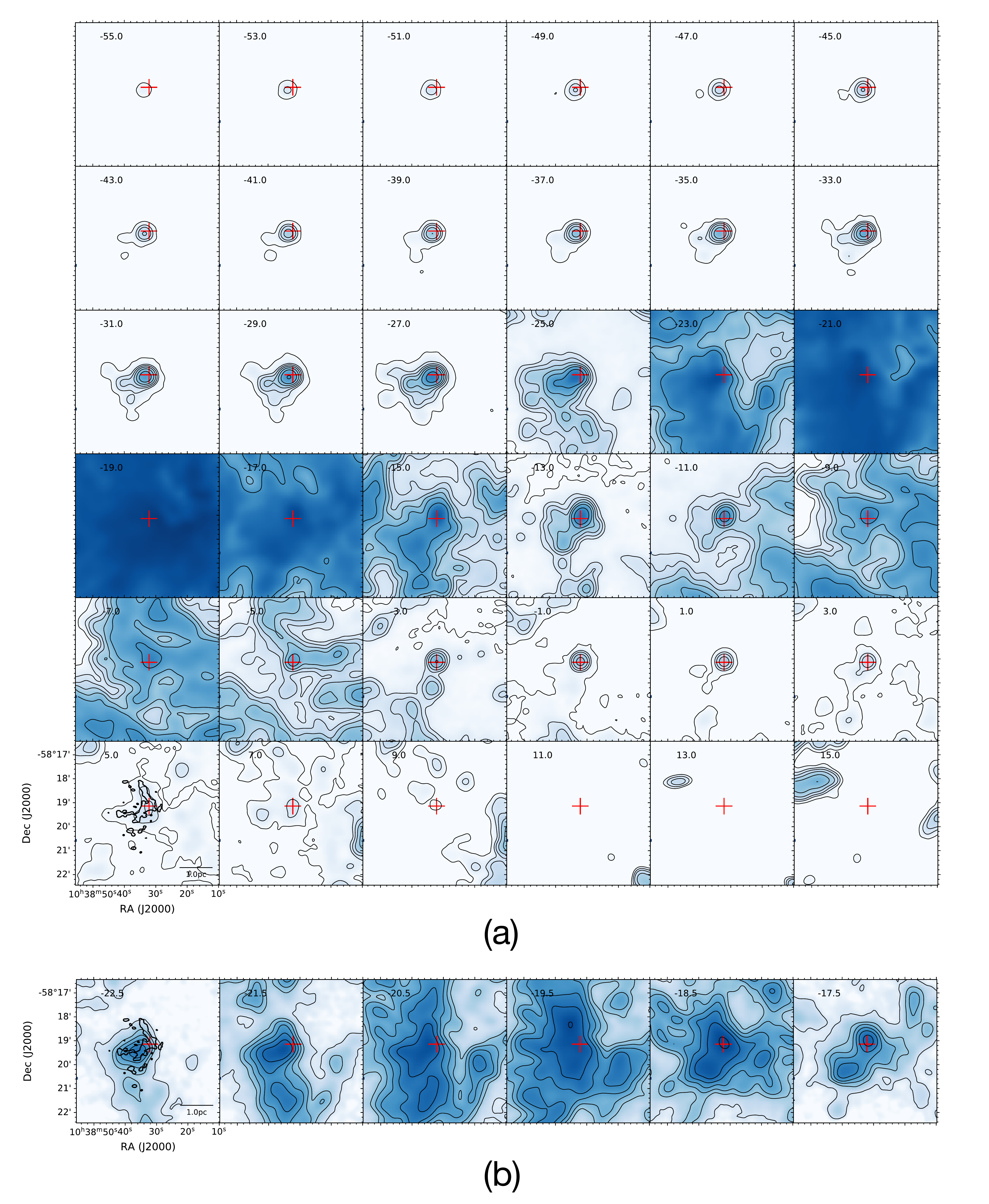}
\caption{
{\it (a)} Channel maps of TP CO(2-1) emission integrated over every 2.0
\kms, as indicated in the upper left of each panel (indicating central
velocity of the range), from -55.0 to +15.0 \kms.  The contour levels
are 1 \jypbm\ \kms $\times$ (1, 5, 10, 20, 40, 80, 200). The red cross in each panel marks the phase center of the
observation (R.A.=10:38:33, decl.=-58:19:22). The thick black contour
in the lower left panel shows the 4$\sigma$ level of the 7-m continuum
emission.  {\it (b)} Channel maps of TP \ceighteeno{}(2-1) emission
integrated over every 1.0 \kms, with ranges from $-$22.5 to $-$17.5
\kms. The contour levels
are 1 \jypbm\ \kms $\times$ (1, 5, 10, 20, 40, 80, 200).
The thick black contour in the left panel shows the 4$\sigma$
level of the 7-m continuum emission.
}
\label{fig:tp_chanmap}
\end{figure*}

To further explore the kinematic structure of the clump, we present
the CO(2-1) channel map from $-55.0\:$\kms\ to 15.0~\kms~ in
\autoref{fig:tp_chanmap}(a), where we have averaged four velocity
channels in each displayed panel. The CO emission is widespread
around the systemic velocity (-23 \kms\ to -17 \kms). Bluewards of the
line center the emission retains extension towards the southeast and
then at the highest blueshifted velocities, e.g., $v\lesssim
-45\:$\kms, appears more concentrated.
The redshifted emission shows more complex structure, including from
emission features already mentioned at around $v=-9\:$\kms, which may
be from an unrelated cloud along the line of sight. However, high
velocity ($\Delta v\gtrsim25\:$\kms) redshifted gas is still seen near
the phase center. These high velocity features, both blue- and
redshifted, are likely to be caused by protostellar outflow activity
from within the G286 star-forming clump.

The clump-averaged spectra could be affected by multiple factors
including collapse, rotation and outflows. To better resolve the kinematics near the systemic velocity where
$^{12}$CO(2-1) is expected to be mostly optically thick, in
\autoref{fig:tp_chanmap}(b) we show the \ceighteeno{}(2-1) channel map
from -23.0 \kms~ to -17.0 \kms. This \ceighteeno{} emission at around
-20~\kms\ is moderately elongated in the North-South direction.
In the central 2\arcmin\ region, the \ceighteeno{}(2-1) at blueshifted
velocities is mostly extended to the southeast, while at the
corresponding redshifted velocities, there is a more complex,
widespread morphology, including some material at northeastern and
southeastern locations.

\subsection{Observations of the 7-m and 12-m arrays}



\begin{figure*}[ht!]
\epsscale{1.2}\plotone{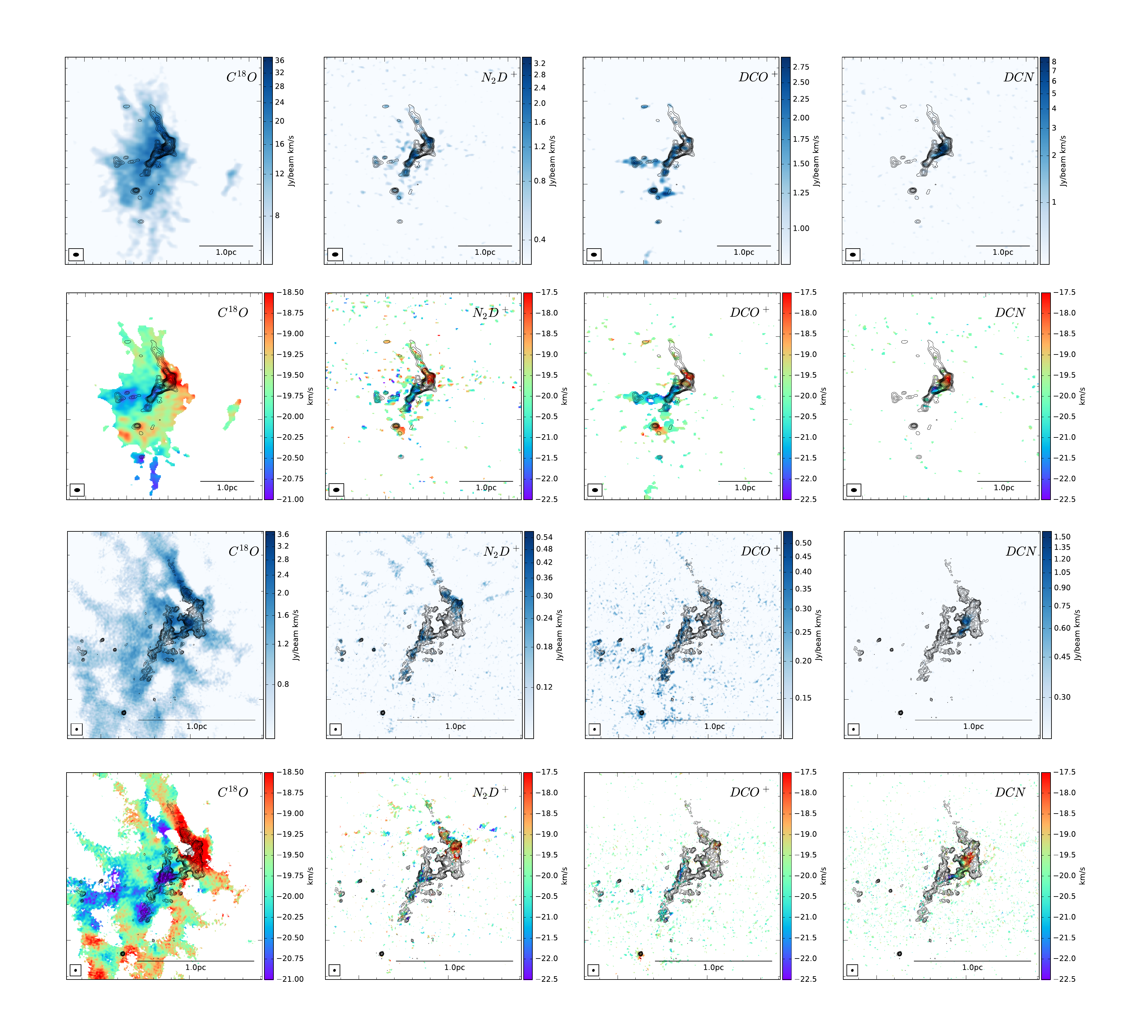}
\caption{
Summary figure for the 7-m and 12-m line observations.  Columns from
left to right show the results of \ceighteeno{}(2-1), \ntwodp{}(3-2),
\dcop{}(3-2) and \dcn{}(3-2), respectively. From top to bottom, the
color scales show the maps of 7-m moment 0, 7-m moment 1, 7-m+12-m
moment 0, 7-m+12-m moment 1. The color bar at the right corner
indicates the flux scale in \jypbm{} for moment 0 maps, and velocity
in \kms~ for moment 1 maps. The black contours illustrates the 1.3~mm
continuum emission for comparision, with the first two rows showing
7-m cotinuum image and last two rows 7-m+12-m image.  }
\label{fig:mom01}
\end{figure*}

\autoref{fig:mom01} presents summary maps of four spectral lines
\ceighteeno{}(2-1), \ntwodp{}(3-2), \dcop{}(3-2) and \dcn{}(3-2) from
left to right overlaid on 1.3~mm continuum image in black contours.
The top two rows show the moment 0 and moment 1 map of the 7-m array
images, respectively.
As shown in the moment 0 map, \ceighteeno\ traces structures
that are more spatially extended than other lines. \ntwodp{} and
\dcop{} are more closely associated with the dust continuum, but their
distributions are slightly different.  \ntwodp{} is mainly detected
towards the NW-SE filament and the southern part of the NE-SW
filament.  Note also that not all the regions with strong dust
continuum have detections of \ntwodp.  In particular, there is a
deficiency of \ntwodp{} emission towards the central brightest
clump. \dcop{} emission, on the other hand, appears slightly more
extended. There is also an E-W filamentary feature to the east of
NW-SE filament. This E-W filament is not seen clearly in continuum
emission, where we only observe a few cores strung out along the EW
direction, but these do appear to be connected by weak diffuse dust
emission seen at a 3-$\sigma$ level. Additionally, \dcop{} is also
detected towards a few positions to the south of NW-SE filament.  The
spatial distribution of \dcn{} emission is dramatically different from
\ntwodp{} and \dcop: it is strongly concentrated towards the clump in
the center, where no detection or only weak detection is seen for
\ntwodp{} and \dcop. \dcn{} emission becomes weaker away from the
center.

The different morphological distributions of the deuterated species
may be due chemical differentiation. In general, \ntwodp{} is known as a
good tracer of cold ($T\lesssim20\:$K), dense gas, where $\rm H_2D^+$
builds up in abundance, but where CO is mostly frozen out on to dust
grains \citep[e.g.,][]{Fontani15}.
Formation of \dcop{} requires both $\rm H_2D^+$ and gas phase CO
\citep[e.g.,][]{Millar89}, which requires a temperature $\lesssim$
30~K but not too cold to cause significant CO freeze-out.
On the other hand, the primary DCN formation mechanisms are thought to
require $\rm CH_2D^+$ instead of $\rm H_2D^+$, which is energetically
favorable up to $\sim$80~K
\citep[e.g.,][]{Millar89,Turner01}. Additionally, sputtering from
grain mantles can also lead to enhancement of DCN abundance in shocked
regions \citep[e.g.,][]{Busquet17}. Hence we would generally expect
more DCN emssion in relatively later evolutionary stages. The
concentrated distribution of DCN, combined with more wide-spread
\ntwodp{} and \dcop{} emission, indicates a sceneario that star
formation, especially more massive, luminous star formation, has taken
place first in the central regions of G286 compared to in the more
extended filaments.

The second row of \autoref{fig:mom01} shows the moment 1 map of the
7-m array images. The \ceighteeno{} moment 1 map reveals redshifted
emission associated with the NE-SW filament and then
continuing to the south of NW-SE filament, while the NW-SE filament
and E-W filament are mainly associated with blueshifted gas.  Other
dense gas tracers show similar velocity patterns as \ceighteeno{},
but with the emission mainly detected towards dense continuum
clumps. In particular, DCN illustrates the blue-red velocity
transition across the central clump in the NW-SE direction.

A zoom-in view of G286 is presented in the third and fourth row of
\autoref{fig:mom01}, illustrating the moment 0 and moment 1 map of
combined 12-m+7-m array image with a resolution of $\sim$ 1.5". The
continuum image reveals a higher level of fragmentation and many
well-defined dense cores, with a typical size of a few thousand AU.
The E-W filament and part of the NE-SW filament are resolved out in
this continuum image. The intensity and velocity distribution of
\ceighteeno{} appears more complicated seen in high resolution. Other dense 
gas tracers like \ntwodp{} still have good
association with continuum at the core scale, and the velocity pattern
is also consistent with that seen in the 7-m image.

\begin{figure*}[ht!]
\epsscale{1.}\plotone{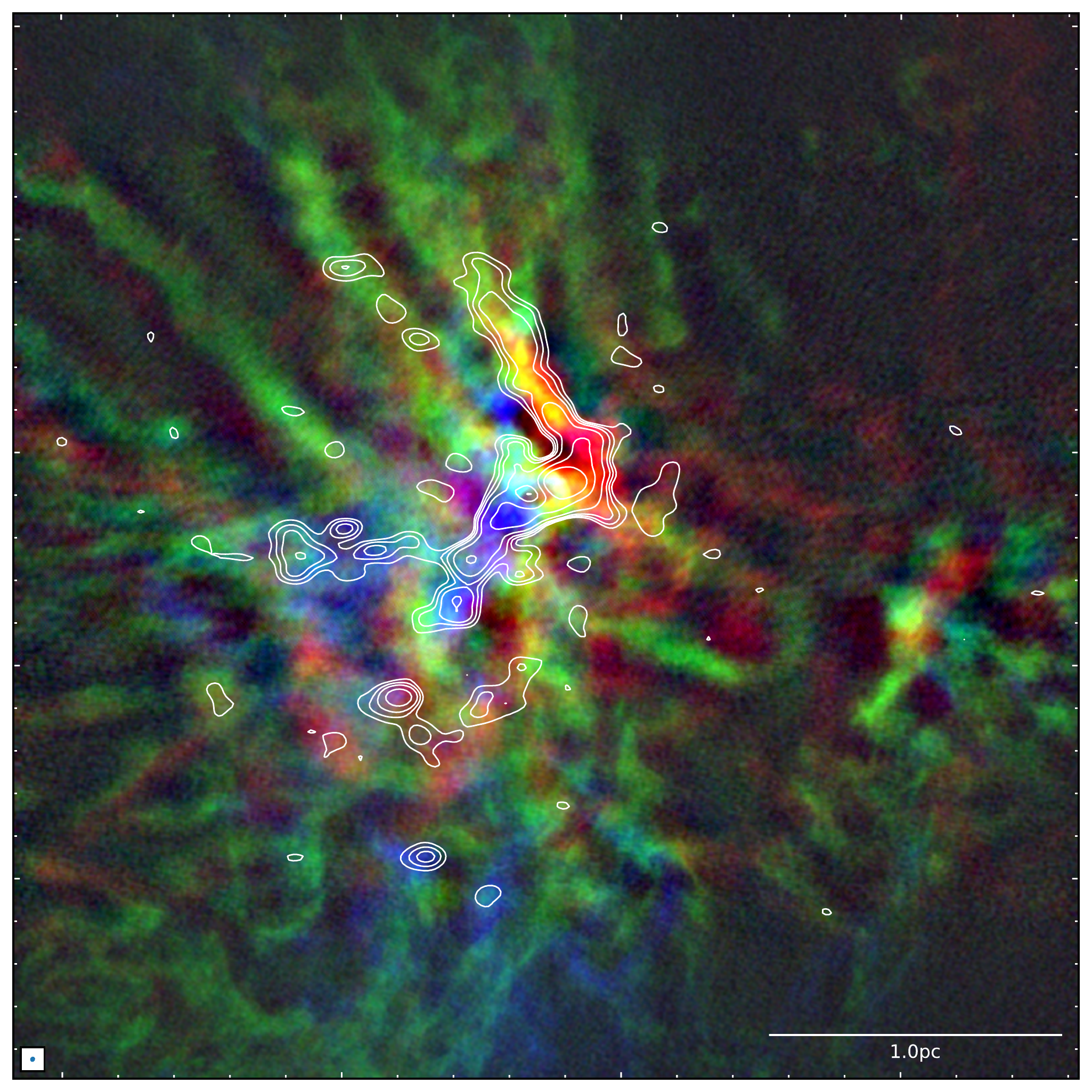}
\caption{
 Three color image constructed with integrated
12-m + 7-m \ceighteeno{}(2-1) emission (-23.0 -- -20.5 \kms~ in blue, -20.5 --
-19.5 \kms~ in green and -19.5 -- -17.0 \kms~ in red). The synthesized beam (1.56\arcsec$\times$1.40\arcsec) is shown in the lower left corner. The 7-m continuum image is shown in white contours
for comparison. The contour levels are 1.7 \mjypbm
$\times$ (3, 6, 10, 20, 50, 100). }
\label{fig:c18o}
\end{figure*}

In \autoref{fig:c18o} we present the 12-m + 7-m \ceighteeno{} image with integrated emission in different velocity intervals shown in different colors, i.e., -23.0 to -20.5 \kms\ in blue, -20.5 to -19.5 \kms\ in green and -19.5 to -17.0 \kms\ in red. Besides the velocity structures seen in \autoref{fig:mom01}, this plot also reveals highly filamentary \ceighteeno{} features around the systemic velocity. These filaments are more spatially extended than the continuum. While some of this morphology may be affected by
artificial sidelobes from imperfect cleaning of the interferometric
data, at least some of the \ceighteeno (2-1) filaments have corresponding detections in the continuum and hence are most likely to be real features.

\section{Filamentary Virial Analysis} \label{sec:fila}

As shown in \autoref{fig:overview}, the millimeter continuum emission
reveals two main filaments: a northern one with a NE-SW orientation and a
southern one with a NW-SE orientation.  Here we perform a filamentary
virial analysis following \citet[][]{Fiege00}.  Since the NE-SW
filament is mostly filtered out at higher resolution, we utilize the
7-m array data (continuum and \ceighteeno) for this section.

\begin{deluxetable}{lccccc}
\tabletypesize{\scriptsize}
\caption{Properties of the NE-SW Filament}
\label{table:fila}
\tablehead{
\colhead{Properties} & \colhead{Strip 1} & \colhead{Strip 2} & \colhead{Strip 3} & \colhead{Strip 4} & \colhead{Total}
}
\startdata
$\overline{\Sigma}_{\rm sed}$ (\gcm)  & 0.25 & 0.17 & 0.13 & 0.12 & 0.17\\
$M_{\rm sed}$ ($M_\odot$)  & 53   & 36   &  27  &  26  &  142  \\
$M_{\rm 1.3mm}$ ($M_\odot$)  & 25   & 17   &  21  &  11  &  74  \\
$m_{\rm sed,f}$ ($M_\odot$ pc$^{-1}$) & 254 & 170 & 130 & 123 & 170 \\   \hline
$\overline{v}_f $ (\kms) & -17.85 & -18.59 & -19.01 & -19.40 & -18.73 \\
$ \sigma_{\rm C^{18}O}$(\kms) & 0.40 & 0.52 & 0.53 & 0.61 & 0.52\tablenotemark{a} \\ 
$ \sigma_{f}$(\kms)  & 0.48 & 0.58 & 0.59 & 0.66 & 0.58 \\
$m_{\rm vir,f}$ ($M_\odot$ pc$^{-1}$) & 106 & 158 & 160 & 204 & 158 \\ \hline
$ m_f/m_{\rm vir,f}$  & 2.39 & 1.08 & 0.81 & 0.60 & 1.08 \\
\enddata
\tablenotetext{a}{For velocity dispersion we take the linear average of 4 strips.}
\end{deluxetable}

\begin{figure*}[ht!]
\epsscale{1.1}\plotone{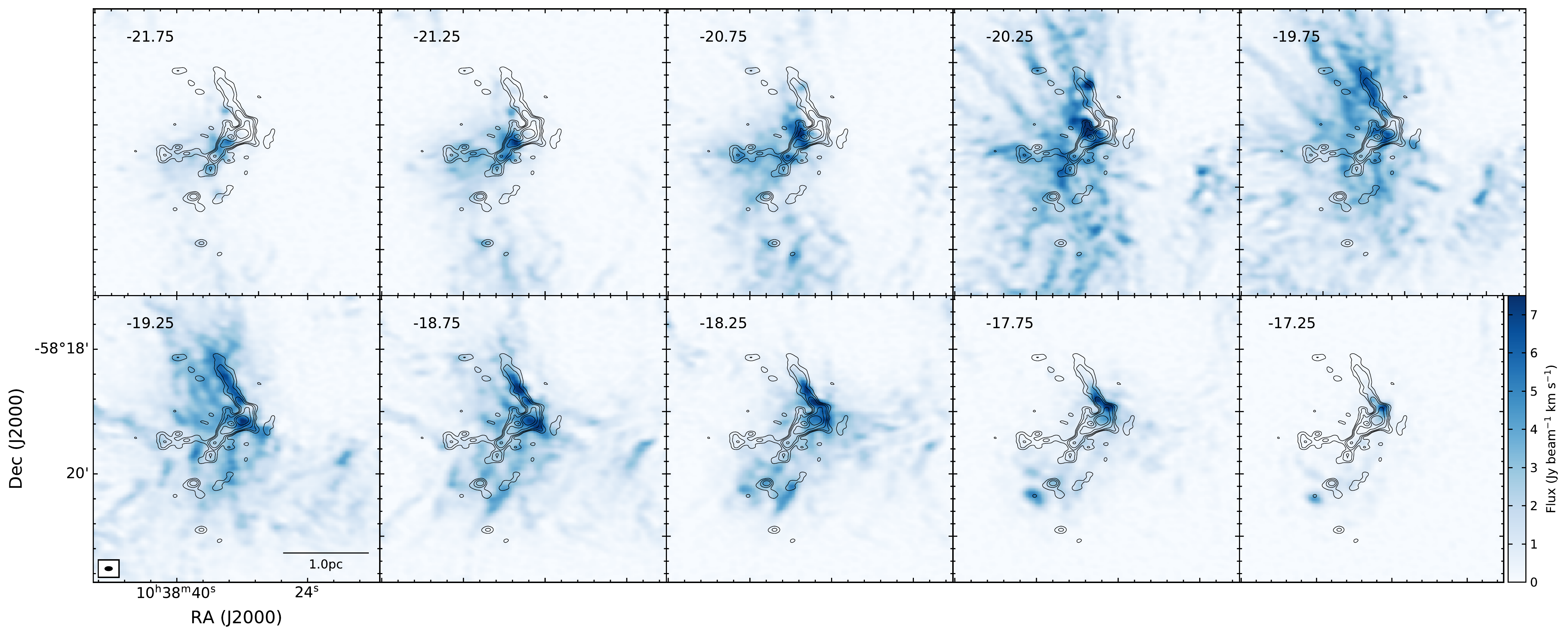}
\caption{
7-m \ceighteeno (2-1) emission integrated over 0.5~\kms\ intervals, as
indicated in the upper left of each panel, from -22.0 to -17.0~\kms.
The black contours show the 7m array 1.3~mm continuum emission.
The contour levels are 1.7  \mjypbm $\times$ (4, 10,
  20, 50, 100).  }
\label{fig:c18o_chanmap}
\end{figure*}

\autoref{fig:c18o_chanmap} illustrates the \ceighteeno(2-1) emission,
integrated over every 0.5 \kms\ from -22.0 to -17.0 \kms.
As in \autoref{fig:c18o}, filamentary structures are seen near the systemic velocity of -20
\kms. 
At least three of the \ceighteeno (2-1) filaments have
corresponding detections in the continuum at a 2$\sigma$ level and
hence are most likely real features, rather than sidelobe artifacts. The most prominent filament is
associated with the NE-SW continuum filament and is clearly seen from
-20.0 \kms{} to -17.0 \kms.  The NW-SE filament appears more complicated in \ceighteeno{}(2-1) and is not
well described as being a coherent \ceighteeno{} filamentary
structure. Therefore we carry out a virial analysis only for the NE-SW
filament.

As shown by \citet{Fiege00}, a pressure-confined, non-rotating,
self-gravitating, filamentary (i.e., length $\gg$ width) magnetized
cloud that is in virial equilibrium satisfies
\begin{equation}
\frac{P_e}{P_f} = 1 - \frac{m_f}{m_{\rm vir,f}}\left(1-\frac{M_f}{|W_f|} \right)
\end{equation}
where $P_f$ is the mean total pressure in the filament, $P_e$ is the
external pressure at its surface, $m_f$ is its mass per unit length,
$m_{\rm vir,f} = 2\sigma_f^2/G$ is its virial mass per unit length,
and $M_f$ and $W_f$ are the gravitational energy and magnetic energy
per unit length, respectively. Here, because of the observational
difficulties of measuring the surface pressure and magnetic fields, we
ignore the surface term and magnetic energy term, i.e., only
considering the balance between gravity and internal pressure support.

\begin{figure*}[ht!]
\epsscale{1.}\plotone{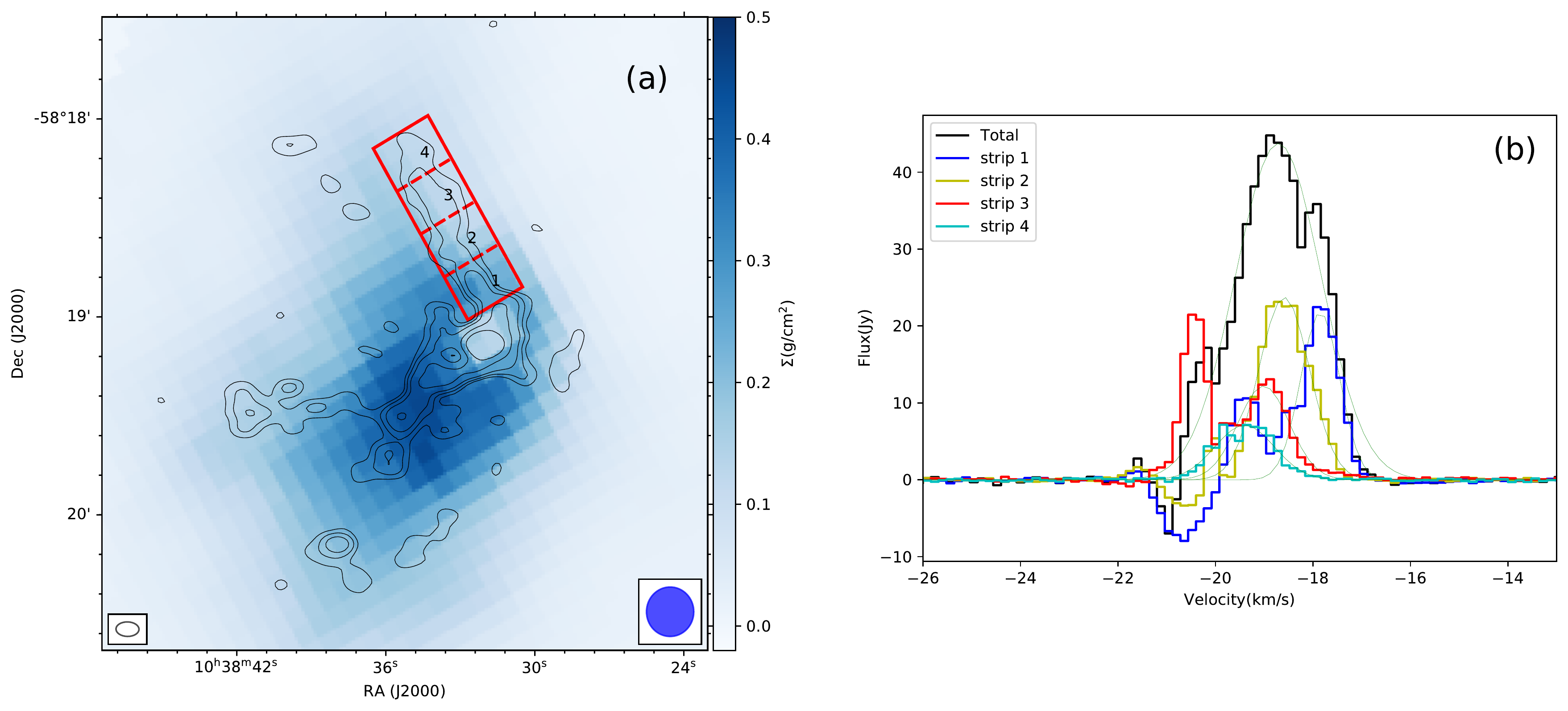}
\caption{
{\it (a)} Column density map made with {\it Herchel} sub-mm continuum
data, overlaid on the 7-m array continuum emission in contours. The
contour levels are 1.7\mjypbm $\times$ (4, 10,
  20, 50, 100). The {\it ALMA} synthesized beam is shown in the lower
left corner, while the resolution of the {\it Herschel}-derived mass
surface density map is shown in the lower right. The red rectangles
dilineate the position of the NE-SW filament and its division into
four strips, numbered 1 to 4 from south to north.  {\it (b)}
\ceighteeno{}(2-1) spectra of the four strips of the NE-SW filament
and the total (see legend). The green lines show primary gaussian
component fits to these spectra.
}
\label{fig:fila}
\end{figure*}

To measure the properties of the filament we show in
\autoref{fig:fila}a a 60\arcsec $\times$ 20\arcsec\ rectangle that
closely encompasses the NE-SW filament, which we use to define the
filament boundary. From the {\it Herschel}-SED-derived mass surface
density map we find average values of $\Sigma_{\rm sed}$ in the strips
ranging from 0.25~$\rm g\:cm^{-2}$ (in Strip 1 that is closest to the
center of G286) to 0.12~$\rm g\:cm^{-2}$ (in Strip 4) (see
\autoref{table:fila}). The mass in each region is then estimated, with
values of between $M_{\rm sed}=$ 26 and 53~$M_\odot$. For comparision,
we also calculate masses from the 1.3~mm continuum flux, assuming a
temperature of 20~K and other dust properties following
\citet{Cheng18}. We find the 1.3~mm-derived mass estimates are about a
factor of two smaller than that measured from the {\it Herschel}-SED
fitting method. Since the {\it ALMA} 7-m array observations only probe
scales up to $\sim$ 19\arcsec, they are likely to be missing some flux
from the filament leading to an underestimation of the masses, and so
here we adopt the {\it Herschel}-SED-derived mass estimates for the
virial analysis.


The 60\arcsec\ length of the filament corresponds to 0.73~pc at an
assumed distance of 2.5~kpc. We assume a 10\% uncertainty in the
distance \citep[e.g.,][]{Barnes10}.
Without direct observational constraints, we further assume the
filament axis is inclined by an angle $i = 60^\circ$ to the line of
sight (90$^\circ$ would be in the plane of the sky). If an inclination
angle of 90 or 30$^\circ$ were to be adopted, then the length
estimates would differ by factors of 1.15 and 0.577,
respectively. Thus the actual the length of the filament is assumed to
be 0.84~pc (or 3/4 of this from the centers of Strip 1 to Strip 4).
Thus the overall mass per unit length of the filament is $m_{\rm
  sed,f}\sim 170\:M_\odot\:{\rm pc}^{-1}$, with Strip 1 having a
higher value of $\sim 250\:M_\odot\:{\rm pc}^{-1}$.

The mean line-of-sight velocity and velocity dispersion of the
filament are measured from the average \ceighteeno{} spectra inside
the rectangular regions.  To reduce contamination from surrounding
ambient gas at the systemic velocity, we utilize the image cube made
with only the 7-m array data (i.e., without feathering with the TP
data), as illustrated in \autoref{fig:fila}b. We perform gaussian
fitting to measure the average centroid velocity $\overline{v}_f$ and
velocity dispersion $\sigma_{\rm C^{18}O}$. 
For strips 1 and 3, two Gaussian components are used for a better
fitting. We then compare the \ceighteeno~ spectra with the spectra of 
denser gas tracers like \dcop~ and found that only one component 
is associated with these tracers. This primary component is 
shown in green lines in \autoref{fig:fila}b and is used for further analysis. 

The values of $\overline{v}_f$ show a steady progression from
$-17.85\:$\kms{} in Strip 1 to $-19.40\:$\kms{} in Strip 4, which
corresponds to an overall velocity gradient of $2.84\:{\rm
  km\:s^{-1}\:pc}^{-1}$ using plane-of-sky projected distance or
$2.46\:{\rm km\:s^{-1}\:pc}^{-1}$ for the assumed $60^\circ$
inclination. We can compare these kinematics to the IRDC filament
studied by \citet{Hernandez11,Hernandez12},
which has a length of 3.77~pc on the sky (4.35~pc for the assumed
$60^\circ$ inclination) and also had its \ceighteeno{}(2-1) emission
analyzed in 4 strips. Here the velocities did not show a steady
progression, but showed differences of about 0.5~\kms\ from strip to
strip, i.e., corresponding to velocity gradients of about $0.53\:{\rm
  km\:s^{-1}\:pc}^{-1}$ in the plane of the sky. The larger and more
systematic velocity gradient shown in the NE-SW filament in G286 may
be the result of acceleration due to infall into the protocluster
potential. Strip 4 has a mean velocity similar to that of the ambient,
larger-scale gas in the region, while Strip 1, closer in projection to
the protocluster center, is redshifted with respect to this
velocity. Thus in this scenario the Strip 4 end of the filament is
closer to us than the protocluster center.

If the velocity change from Strip 4 to Strip 1, i.e., $+1.55\:{\rm
  km\:s}^{-1}$, is due to infall in the protocluster potential, then
we can use this information to constrain the mass of the
protocluster. Assuming an uniform distribution of matter in a spherical
protocluster clump of radius $L$, the change in potential from the
edge to the center is $GM/(2L)$. If material starts at rest at radius
$L$, i.e., the Strip 4 position, and then accelerates to velocity
$v_1$, of which we observe $v_1\:{\rm cos}i$, then the mass inside
radius $L$ is
\begin{equation}
M = \frac{232}{{\rm cos}^2i\:{\rm sin}\:i} \left(\frac{v_{\rm 1,obs}}{\rm km/s}\right)^2\left(\frac{L_{\rm obs}}{\rm pc}\right)\:M_\odot.
\end{equation}
For an observed length $L_{\rm obs}$ from the center of Strip 4 to the center of Strip
1 of 0.55~pc (i.e., 3/4 of 0.73~pc) and a line of sight velocity difference of 1.55 \kms~, we thus estimate the dynamical mass to be $1410\:M_\odot$, assuming $i = 60^{\circ}$. If an inclination angle of 30$^{\circ}$ or 70$^{\circ}$ is adopted, the mass would be 814 or $2780\:M_\odot$, respectively. This estimation based on filament infall kinematics is broadly consistent with that derived from {\it Herschel}-SED fitting, i.e., $\sim2900\:M_\odot$ for the region defined by the larger ellipse aperture in \autoref{fig:overview}. Note, for this SED-based method, we expect $\sim$50\% uncertainty in the mass estimation due to dust opacity and distance uncertainties.

Considering the internal dynamics of the filament, in order to account
for support against gravity from both thermal and non-thermal motions
of the gas, we subtract the thermal component of broadening of the
\ceighteeno{}(2-1) line from the measured velocity dispersion (in
quadrature, assuming a temperature of 20~K) and add back the sound
speed to obtain the total
1D velocity dispersion, $\sigma_f$, i.e.,
\begin{equation}
\sigma_{f} =  \left(\sigma_{\rm nth}^2+\sigma_{\rm th}^2\right)^{1/2} \\
             =  \left(\sigma_{\rm C^{18}O}^2-\frac{k_BT}{\mu_{\rm C^{18}O} m_p}+\frac{k_BT}{\mu_p m_p}\right)^{1/2}
\end{equation}
where $\mu_p$ = 2.33 is
the mean molecular weight assuming $n_{\rm He}=0.1 n_{\rm H}$ and $\mu_{\rm C^{18}O}$ is the molecular 
weight of \ceighteeno.
We have then carried out a virial 
analysis for each of the four strips (see
\autoref{table:fila}). Note, for Strips 1 and 3 we fit the
spectra with two gaussian components and utilize the component that is
more clearly associated with the filament. For example, in Strip 3,
the velocity component near -20.5 \kms~ is contributed by another gas
clump to the north-west of the filament.

The values of $m_f/m_{\rm vir,f}$ of the four strips range from 0.60 to
2.39. Given the systematic uncertainties in measuring the masses and
lengths of the structures that combine to be at least $\sim50\%$,
these values are consistent with the filament being in approximate
virial equilibrium, even without accounting for surface pressure and
magnetic support terms. We also note that the values of
$m_f/m_{\rm vir,f}$ grow, i.e., becoming less gravitationally bound, as
one progresses from Strip 4 to Strip 1. This may indicate that infall
motions and/or tidal forces towards the center of the protocluster act
to stabilize the filament.




\section{Kinematic properties of the dense core sample} \label{sec:kine}

\begin{figure*}[ht!]
\epsscale{1.}\plotone{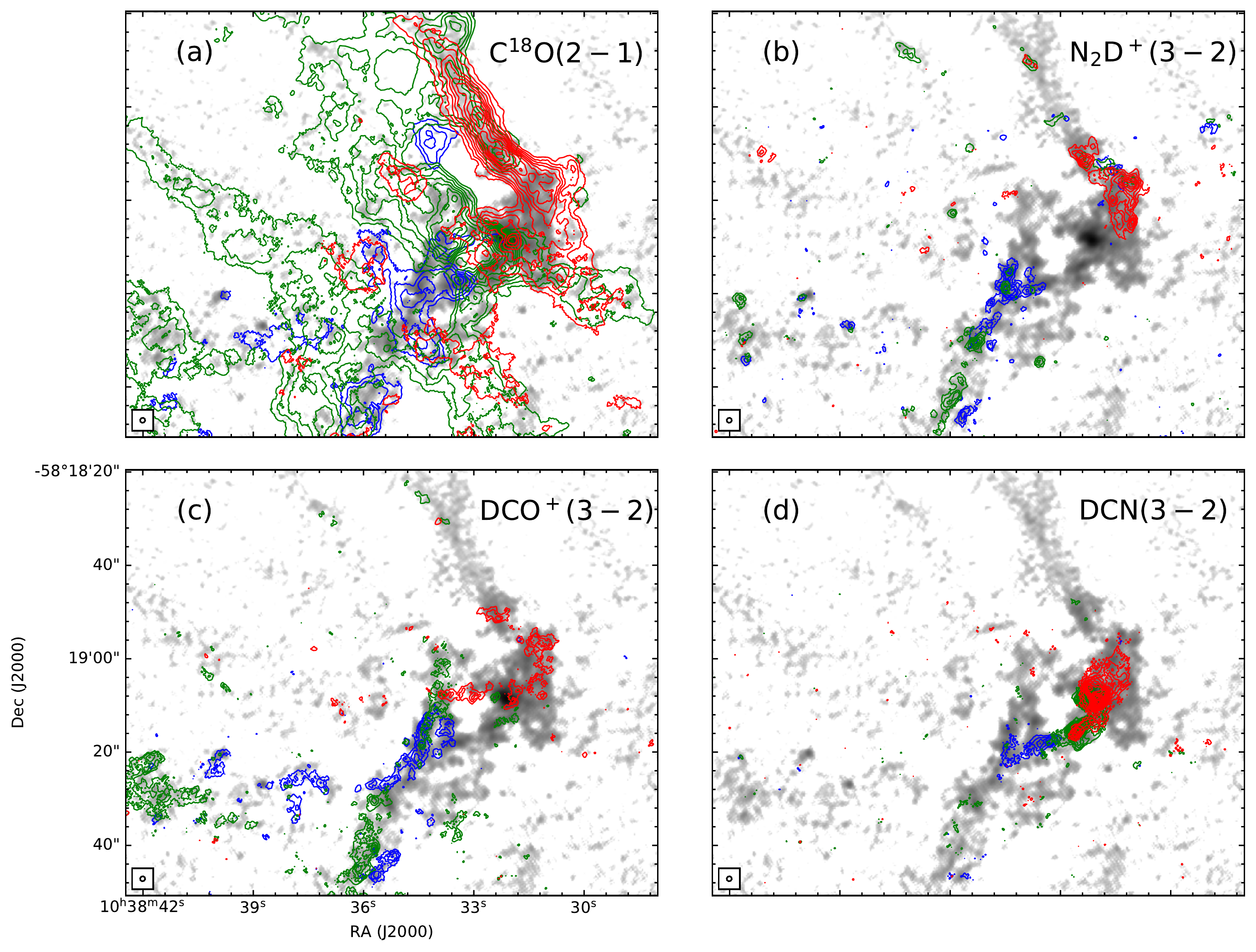}
\caption{
{\it (a)} \ceighteeno{} integrated intensity map using combined 7-m +
12-m array data. Red, green and blue contours show emission integrated
from -23 to -21 \kms, -21 to 19 \kms~ and -19 to -17 \kms,
respectively. The contours start from 4$\sigma$ in step of 2$\sigma$,
with $\sigma$ = 0.1 \jypbm $\cdot$ \kms. The grey scale image is the
1.0\arcsec resolution 7-m + 12-m array combined 1.3~mm continuum
image. {\it (b)} Same as panel (a), but for \ntwodp(3-2). The contours
start from 4$\sigma$ in step of 2$\sigma$, with $\sigma$ = 0.025 \jypbm
$\cdot$ \kms.  {\it (c)} Same as panel (a), but for \dcop(3-2). The
contours start from 4$\sigma$ in step of 2$\sigma$, with $\sigma$ =
0.03 \jypbm $\cdot$ \kms.  {\it (d)} Same as panel (a), but for
\dcn(3-2). The contours start from 4$\sigma$ in step of 2$\sigma$,
with $\sigma$ = 0.03 \jypbm $\cdot$ \kms.  }
\label{fig:12m_map}
\end{figure*}

\begin{figure}[ht!]
\epsscale{0.8}\plotone{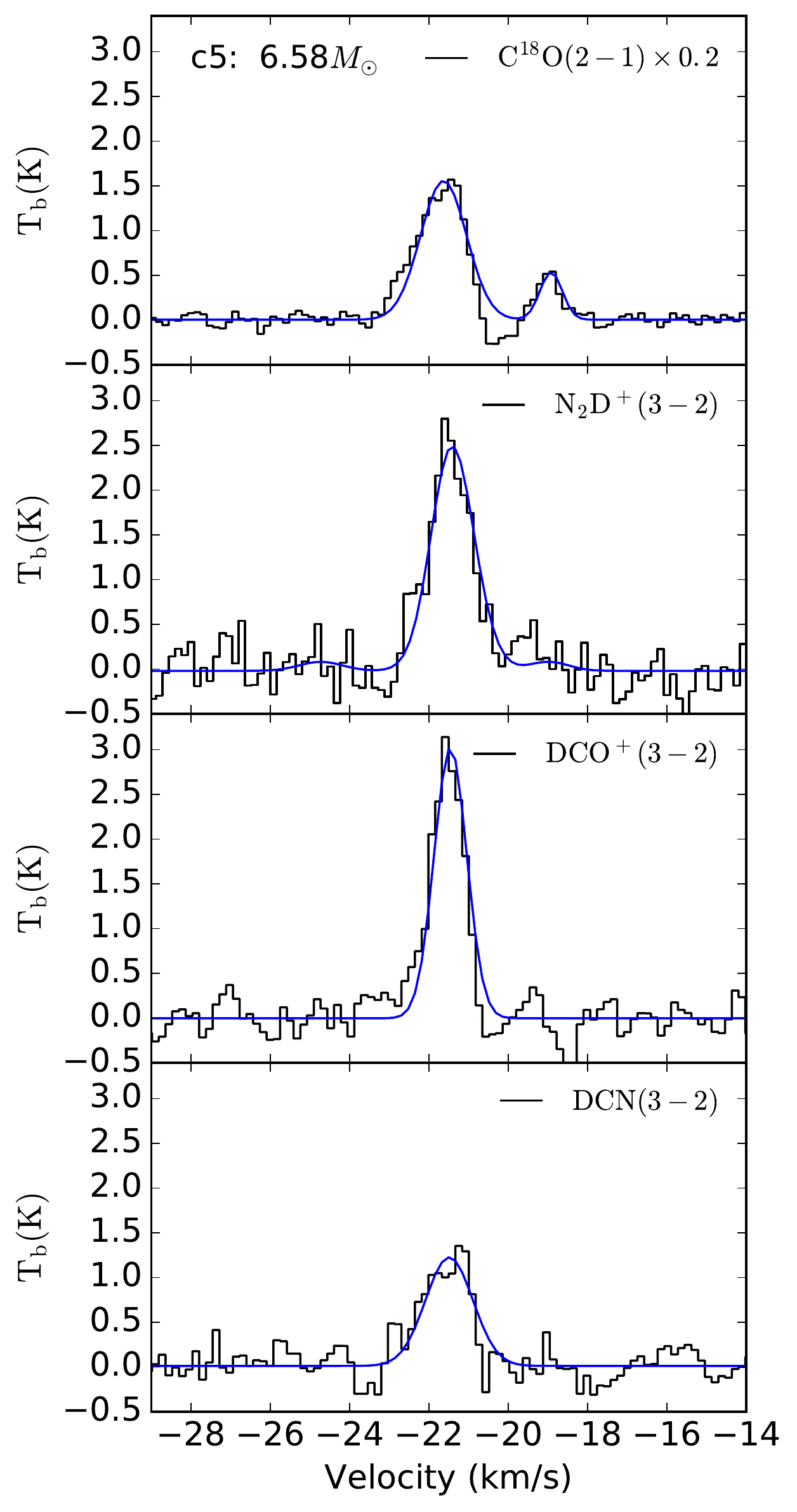}
\caption{
An example of the spectral line fitting for G286c5. Here we use one
gaussian component to fit the spectra of \ntwodp(3-2), \dcop{}(3-2)
and \dcn(3-2), and two components for \ceighteeno{}(2-1).  }
\label{fig:line_example}
\end{figure}

\citet{Cheng18} analysed the mass distribution of dense cores towards
the central region of G286 (about 2.2\arcmin$\times$1.5\arcmin), where
the uv coverage of the observation allows imaging with $\sim$1\arcsec~
resolution.  Here we carry out a kinematic follow-up study on the
dense core sample in this region.



\autoref{fig:12m_map} shows the integrated intensity map of
\ceighteeno{}(2-1), \ntwodp(3-2), \dcop{}(3-2) and \dcn{}(3-2) in the
central region, with three velocity ranges shown in different
colors. This map is similar to the 12-m + 7-m moment maps in
\autoref{fig:mom01}, but emphasizes relatively weaker features that
might be missing in \autoref{fig:mom01} due to higher noise resulting
from its wider velocity range.
Most cores in this region have significant detection from at least one
of the three dense gas tracers: \ntwodp(3-2), \dcop{}(3-2) and
\dcn(3-2), and this allows us to measure the centroid velocity and
velocity dispersion for each dense core.

\subsection{Review of the core sample based on dust continuum emission}

\citet{Cheng18} reported different numbers of identified cores,
ranging from 60 to 125, depending on the detection algorithm and
parameter choices of these algorithms. Here we adopt the fiducial
dendrogram identified core sample with a base threshold of 4$\sigma$,
a delta threshold of 1$\sigma$, along with a minimum area of half a
synthesized beam size. This parameter combination yields 76 cores.

In \autoref{table:cores} we list the properties of the dense core
sample.  The cores are here named as G286c1, G286c2, etc., with the
numbering order from highest to lowest core mass. The masses are
estimated to range from 0.19~$M_\odot$ to 80~$M_\odot$, assuming a
constant temperature of 20~K for each core (see \citet{Cheng18} for
more details). The radius is evaluated as $R_c$ = $\sqrt{A/\pi}$,
where $A$ is the projected area of the core. The median radius is
0.011~pc, similar to the spatial resolution ($\sim$1\arcsec, 2500~AU),
indicating many cores are not well resolved. Note that we adopt the
core area returned by {\it Dendrogram}, which is defined with an
isophotal boundary at a certain flux level, i.e., the level where two
cores merge together or the 4$\sigma$ flux threshold for isolated
cores. So the core area or radius could be underestimated in a crowded
field.

We then evaluate the mean mass surface density of the cores as
$\Sigma_c \equiv M/A$. The median mass surface density of our sample
is $\sim$ 0.65~\gcm\ and all the cores have values $\gtrsim0.4\:$\gcm.
We also evaluate the mean H nuclei number density in the cores,
$n_{{\rm H},c} \equiv M_c/(\mu_{\rm H} V )$, where $\mu_{\rm H}$ =
1.4$m_{\rm H}$ is the mean mass per H assuming $n_{\rm He}$ =
0.1~$n_{\rm H}$ and $V = 4\pi R_c^3/3$.  The mean value of
log$_{10}$($n_{{\rm H},c}/{\rm cm}^{-3}$) is 6.88, with a standard deviation
of 0.24.

\subsection{Spectral fitting}

We extract the average \ceighteeno(2-1), \ntwodp(3-2), \dcop(3-2) and
\dcn{}(3-2) spectra of each core, which are shown in
\autoref{fig:cspec1} and \autoref{fig:cspec2}.
Among the four tracers
\ceighteeno{} is the strongest for almost all the cores, and sometimes
the \ceighteeno{} profiles can be complex. Other lines are relatively
weak and only detected for part of the core sample.

To measure the centroid velocity and velocity dispersion of each core
we only fit spectra with well defined profiles, i.e., those with a
peak greater than a certain threshold value.  Here we adopt a
4$\sigma$ criterion for this threshold value.  Since the noise levels
of the average spectra vary for different cores (depending on the
pixel numbers in the core, etc.), we estimate the rms noise separately
for each core and each line using the signal-free channels. This
signal to noise criterion gives 74 cores detected in
\ceighteeno{}(2-1)(97\%), 27 in \ntwodp{}(3-2)(36\%), 45 in
\dcop{}(3-2)(59\%) and 29 in \dcn{}(3-2)(38\%).
We also checked the single pixel spectra at the continuum peak of each
core and found that the vast majority have similar line profiles as
the averaged spectra, but the signal to noise ratios are usually
lower, so we proceed with our analysis using the core-averaged spectra.

We characterize the \ceighteeno{}(2-1) spectra with 1-d gaussian
fitting using the {\it curve\_fit} function in the {\it
  Scipy.optimize} python module. Most cores can be well described with
a single gaussian component. In general, we expect that
\ceighteeno{}(2-1) traces lower density envelope gas surrounding the
dense core and thus could be more affected by multiple components
along the line of sight. In 31 cores where a spectrum has more complex
profiles and hence can not be well approximated by a single gaussian,
we allow for a second gaussian component.

For the \dcop{}(3-2) and \dcn{}(3-2) lines we also perform the
gaussian fitting with {\it curve\_fit} function.  For the \ntwodp{}(3-2)
line, to account for the full blended hyperfine components, we use the
hyperfine line fitting routine in {\it pyspeckit} \citep{Ginsburg11},
with the relative frequencies and optical depths for \ntwodp{} taken
from \citet{Dore04} and \citet{Pagani09}.  These dense gas tracers are
usually well described with one gaussian
component. \autoref{fig:line_example} shows a example of the line
fitting.  In particular, in one case (i.e., G286c3), two separate
components were clearly required for a good fit for \ntwodp{} and
\dcop.  These two components are mostly likely to belong to two
separate entities that are not resolved in their continuum emission.


\begin{figure*}[ht!]
\epsscale{0.8}\plotone{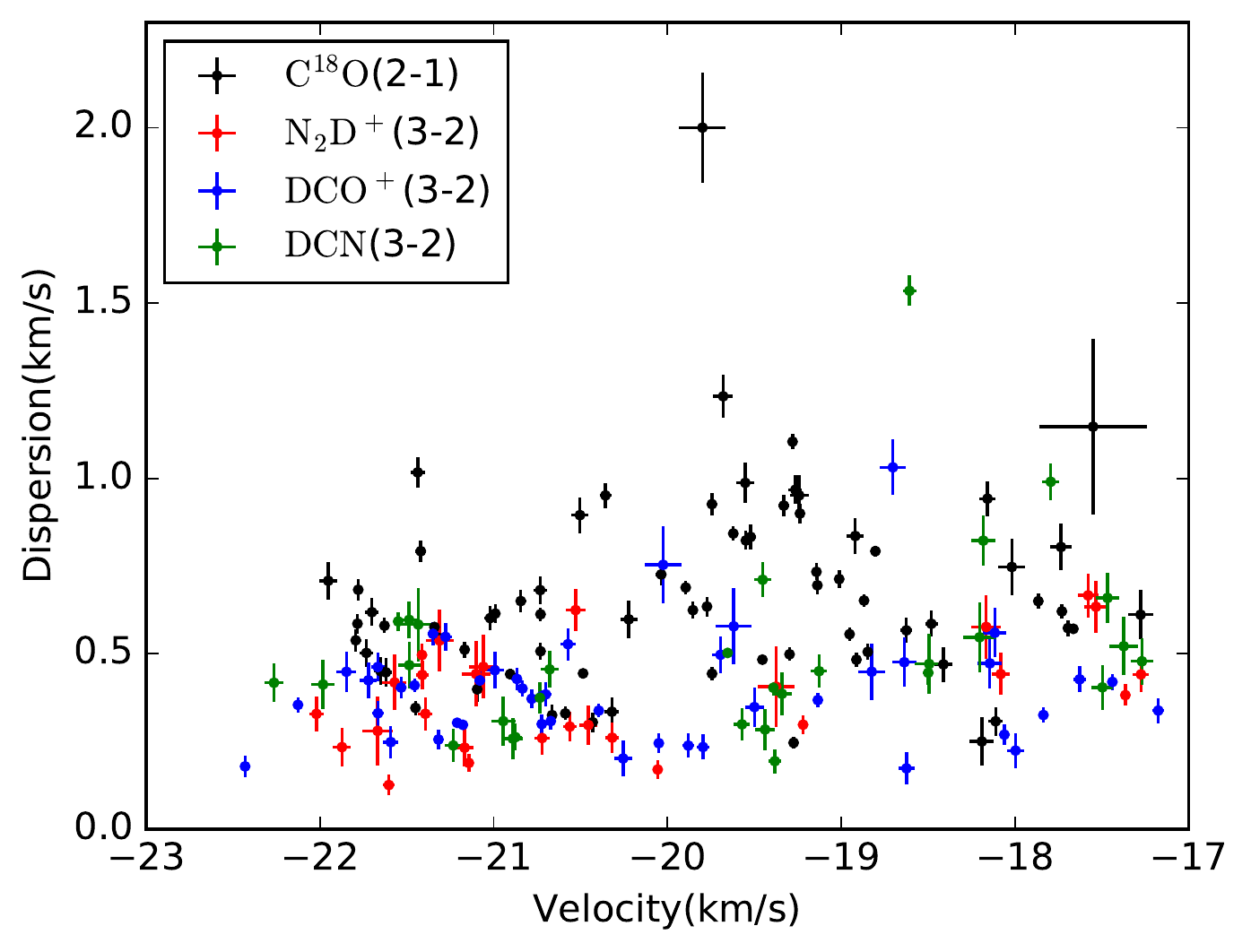}
\caption{
The centroid core velocity and velocity dispersion of each core
measured with \ceighteeno{}(2-1), \ntwodp{}(3-2), \dcop{}(3-2) and \dcn{}(3-2).}
\label{fig:stat1}
\end{figure*}

\subsection{Comments on individual cores}\label{sec:comments}

G286c1: This is the most massive core in G286, with a mass of
80 $M_\odot$ and an equivalent radius of 0.036~pc.  G286c1 is
associated with strong infrared emission and a wide angle bipolar CO
outflow (Cheng et al. in prep.), and hence it is already in a
relatively evolved protostellar stage.  If we adopt a higher
temperature such as 70~K, typical of massive protostellar sources
\citep[e.g.,][]{Zhang18}, then its mass would be $\sim
20\:M_{\odot}$. G286c1 is not detected in \ntwodp{}(3-2), but we see
broad line profiles from \dcop{}(3-2), \dcn{}(3-2) and
\ceighteeno{}(2-1).  In particular, there is very strong \dcn{}(3-2)
emission from -22 to -15~\kms~, which is even broader than
\ceighteeno{}.  Our high resolution ALMA observation in Cycle 5 has
revealed further fragmentation and substructures in G286c1
(Cheng et al., in prep.).  Here, we still use a
one-gaussian component to model the spectral lines of G286c1, and the
resulting fitting parameters should be treated more cautiously as reflecting the
average properties of the core.

G286c3: This is also a massive core, with $\sim 12\:M_\odot$.  We have
detected \ceighteeno{}(2-1), \ntwodp{}(3-2), \dcop{}(3-2) and
\dcn{}(3-2) towards G286c3. Interestingly, these spectra of
\ceighteeno{}, \ntwodp{} and \dcop{} all exhibit a double-peak
profile, with one peak centered around -19.5~\kms~ and another at
$\sim$ -18\kms, though \dcn{} is only detected in one velocity
component.  Since we expect the deuteratated species such as \ntwodp{}
and \dcop{} to be optically thin, these line profiles are more likely
to be contributed by two separate entities inside G286c3 instead of a
central dip caused by self-absorption.  A detailed inspection from the
continuum also reveals that G286c3 is very elongated in the NE-SW
direction. Thus it is possible that there are further
sub-fragmentations in G286c3 that are not identified by our fiducial
dendrogram algorithm: e.g., there could be two cores overlapping along
the line of sight. Here we use two-component gaussian fitting to model
the spectrum of \ceighteeno, \ntwodp{} and \dcop, and treat them as
two individual cores (i.e., two data points per line in
\autoref{fig:stat1}). We split the mass of G286c3 assuming that 
the mass of each component is proportional to the \ceighteeno{} flux 
for relavent analysis.

G286c4: This core has an estimated mass of $\sim 9\:M_\odot$. There is
no \ntwodp{} detection, but we see very strong \ceighteeno{}, \dcop{}
and \dcn{} emission. \dcn{}(3-2) has a very strong peak centered at
-19.5 \kms~, similar to \ceighteeno{} and \dcop{}. Additionally,
there are two secondary peaks at around velocity -16 \kms~ and
-23 \kms. These may be a real features resulting from unresolved
condensations, or more dynamical activities like outflows, but we are
unsure about its origin with the current information. Here for \dcn{} we
only fit the central major velocity component that is consistent with
other tracers.

G286c8, G286c20 and G286c41: These are special in terms of their
\dcop{}(3-2) spectra. All three cores have a \dcop{} peak around
-18 \kms. For G286c20 and G286c41, \dcop{} has a large velocity offset
($\sim$1 \kms) compared with other tracers, like \ceighteeno{}. For
G286c8, this offset is even larger ($\sim$3 \kms) and there is another
obvious \dcop{} peak around -21 \kms, similar with the peaks of
\ceighteeno{} and \dcn{} lines.  A possible explanation is that G286c8
has a core velocity around -21 \kms, as traced by multiple tracers,
while the \dcop{} feature around -18 \kms~ is not associated with the
dense core.  From the continuum map we find that all these three
cores are close together and lie on a filamentary
feature that is only seen in \dcop{}.  This filamentary feature is
clear in the \dcop{} channel map and does not appear to be associated
with dense dust continuum. Hence we exclude this \dcop{} velocity
component near -18 \kms~ for G286c8, G286c20 and G286c41 in our
analysis.

\subsection{Line parameters of different tracers} \label{sec:para}

The best-fit parameters of centroid velocity and velocity
dispersion are displayed along with the spectral lines in
\autoref{fig:cspec1} and \autoref{fig:cspec2}. 
\autoref{fig:stat1}
illustrates the distribution of these parameters, along with their individual uncertainties.  
As can be seen, the centroid velocities
range from -22.5 to -17.0 \kms~ and there is a modest
clustering near -21.5~\kms.  The velocity dispersions range from
0.1 to 1.0~\kms~ for deuterated species, while those of
\ceighteeno{} are systematically larger. \ntwodp{} and \dcop{} usually
give smaller velocity dispersions, with a median value of 0.35
and 0.36~\kms, respectively.  \dcn{}-measured dispersions are larger, with a median value of 0.43 \kms.  Centroid velocity uncertainties range from 0.01 to 0.08~\kms, while velocity dispersion uncertainties vary from 1\% to 20\%, with a
few cases $\gtrsim$30\%, depending on the signal to noise ratio and
shape of the line profiles.

\begin{figure*}[ht!]
\epsscale{1.1}\plotone{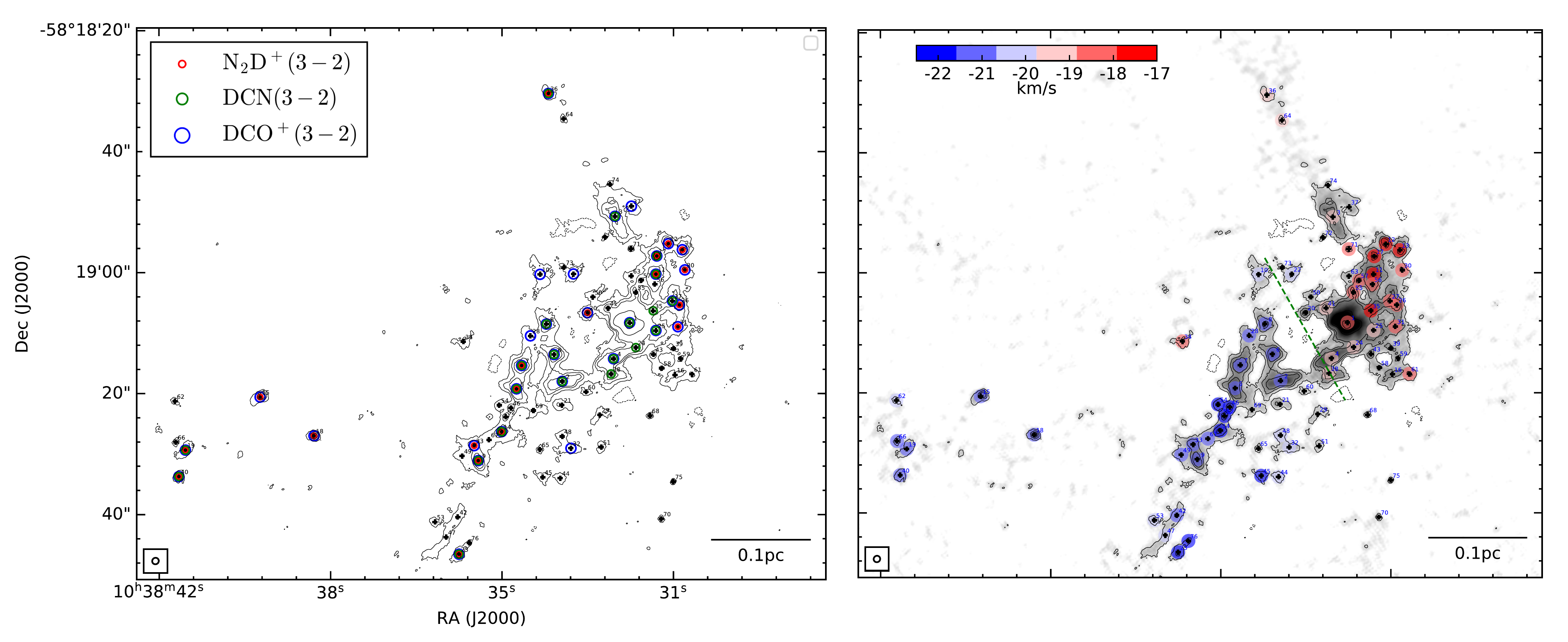}
\caption{
{\it Left:} Line detection status for each core overlaid on the
1.0\arcsec~ resolution 1.3~mm continuum image in contours. 
The black crosses denote positions of cores identified via 1.3~mm continuum by \citet{Cheng18}.
A red circle indicates a detection of \ntwodp(3-2); a blue circle of \dcop{}(3-2); and 
a green circle of \dcn(3-2).
{\it Right:} Core velocity map overlaid on the 1.0\arcsec~ resolution 1.3~mm continuum image shown in contours and greyscale. The core velocity is determined
by averaging the results from \ntwodp, \dcop{} and 
\dcn (see text).
The colored circles indicate the velocity of the 54 dense core
that are detected in at least one deuterated tracer. 
Overall the velocity distribution can be 
described as being composed of two velocity groups that are spatially distinct, with an approximate boundary shown by the dashed
green line.
}
\label{fig:core_vel}
\end{figure*}

\autoref{fig:core_vel} illustrates the line detection situation of the 
core sample, with different colored circles denoting detections
in \ntwodp(3-2), \dcop{}(3-2) and \dcn{}(3-2). \ceighteeno{}(2-1) is detected for
almost all the cores (except c68, c75) and hence is not shown here. 
As already apparent in \autoref{fig:12m_map}, \dcn{}(3-2) is mostly detected in 
the central region, while \ntwodp{}(3-2) and \dcop{}(3-2)-detected
cores are more widespread. 
Overall we have 54 cores that are detected in at least one of these three 
dense gas tracers. In particular, all the cores with \ntwodp{} detection 
also have strong \dcop{} emission. 

\begin{figure*}[ht!]
\epsscale{1.1}\plotone{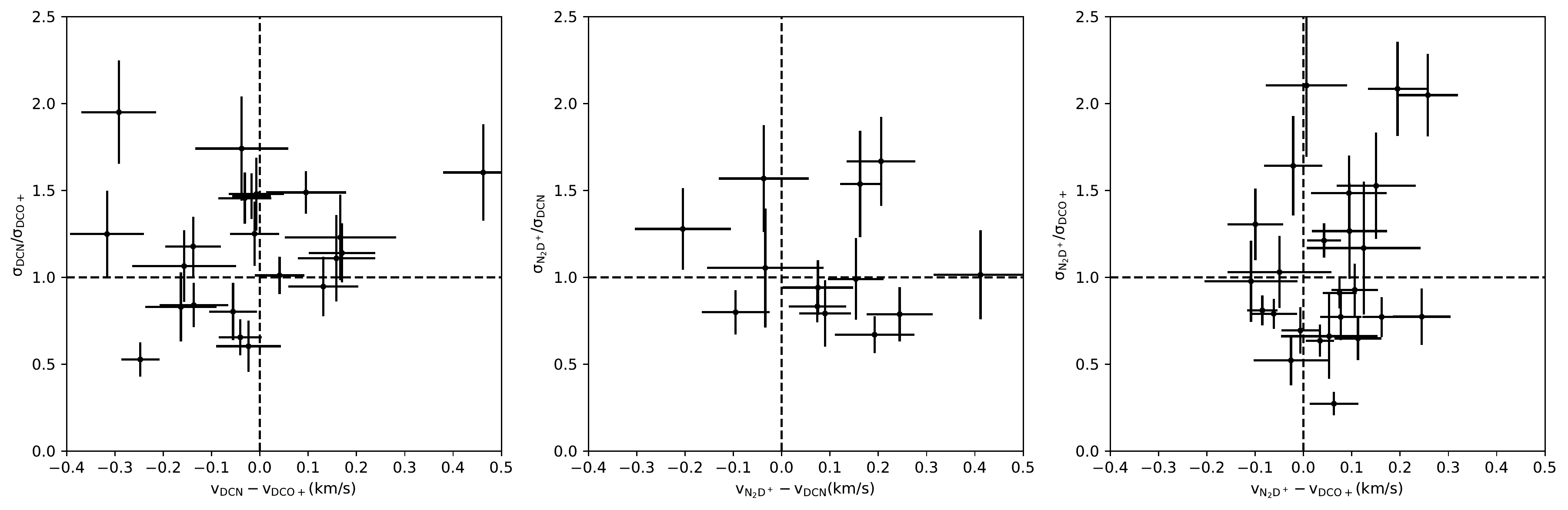}
\caption{
Core centroid velocity differences and relative velocity dispersions as measured from \ntwodp(3-2), \dcop{}(3-2) and DCN(3-2).}
\label{fig:stat2}
\end{figure*}

The cores that are detected in more than one line are of particular
interest, since differences in fitted parameters could be a reflection
of chemical differentiation.
There are 14 cores that are detected in all three lines; 
26 cores that are detected in both \ntwodp{} and \dcop; 
14 in both \ntwodp{} and \dcn; and 21 in both \dcop{}
and \dcn. \autoref{fig:stat2} illustrates the differences in fitted 
parameters of these species when commonly detected. 
From this figure we see that there is no 
significant offset in centroid velocity or velocity dispersion as derived from the different species. This similarity in velocity distributions is expected if these species are tracing the same molecular gas material.

For the centroid velocities, the median offsets between \ntwodp{} and
\dcop, \ntwodp{} and \dcn, and \dcop{} and \dcn{} are 0.07, 0.09, 0.03
\kms, respectively.  The sampling error of the velocity offset
distribution due to the finite number of cores is estimated to be
about 0.04~\kms, so these offsets are not very significant.

The 1d velocity dispersion $\sigma$ are generally consistent among 
different tracers within a factor of 2. The median values of 
$\sigma_{\rm DCN}/\sigma_{\rm DCO^+}$, $\sigma_{\rm N_2D^+}/\sigma_{\rm DCN}$
and $\sigma_{\rm N_2D^+}/\sigma_{\rm DCO^+}$ are 1.16, 0.99 and 0.95, 
respectively. The observed scatter is consistent with the fitting uncertainties.





We next compare dense core centroid velocities with the
larger-scale gas reservoir (or envelope) traced by \ceighteeno. 
Previous studies in relatively low-mass environments have shown that 
cores mostly have subsonic core-to-envelope
motions \citep[e.g.,][]{Walsh04,Walsh07,Kirk07,Walker-Smith13}. 
Our work here provides a measure of core-to-envelope 
motions within a more massive protocluster.
Additionally, most previous works measured the centroid velocity offset between \ceighteeno{} and \ntwohp{}.  Here we have observations of lines from deuterated species like
\ntwodp, \dcop{} and \dcn, which should be better tracers of localized dense cores, rather than more extended filaments, and usually not affected by multiple velocity 
components that may complicate the interpretation \citep[e.g.,][]{Henshaw14,Ragan15}.

\begin{figure*}[ht!]
\epsscale{0.8}\plotone{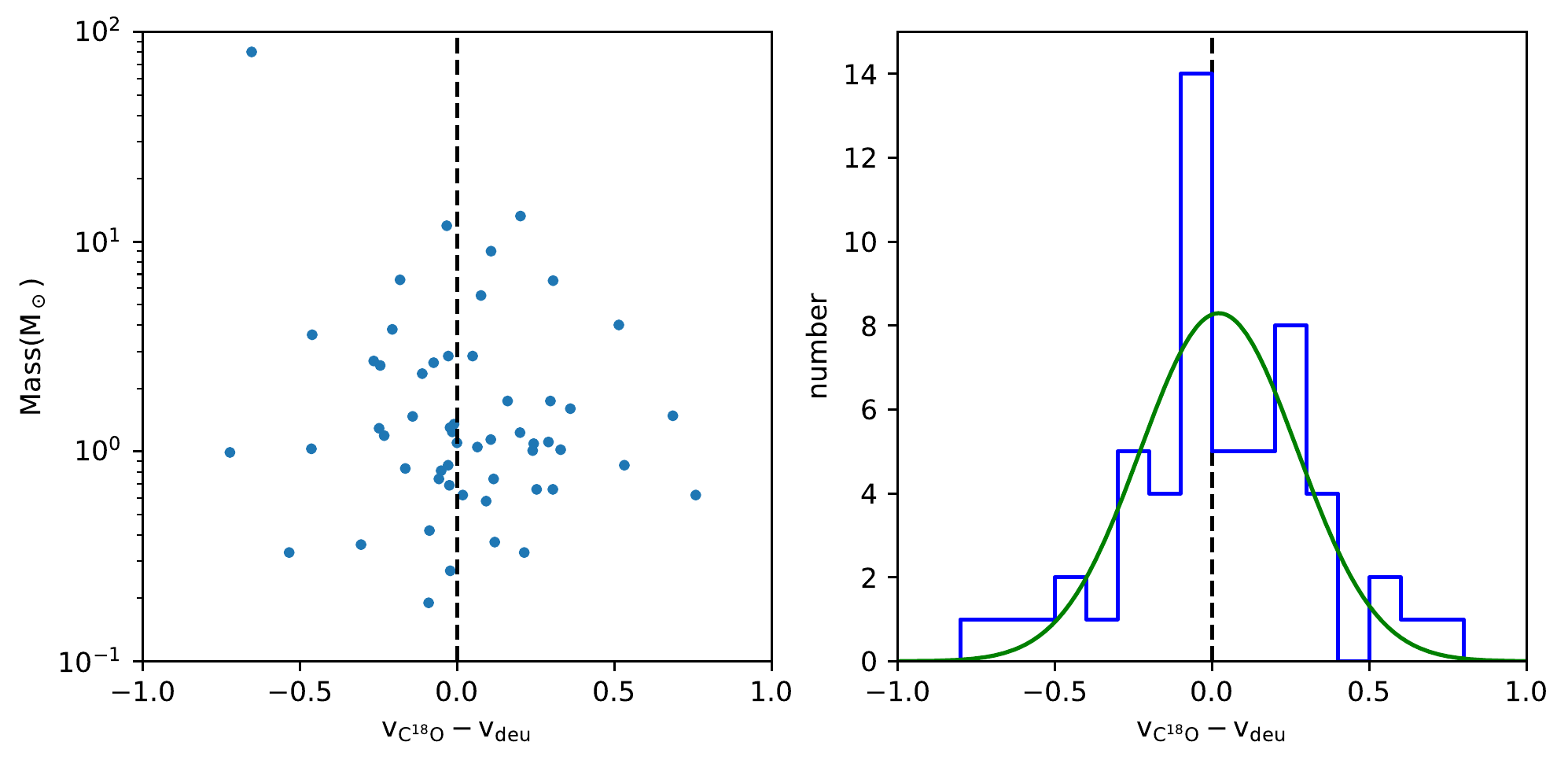}
\caption{{\it Left:} Distribution of differences between the velocity of dense cores
(determined with deuterated tracers) and centroid velocity of \ceighteeno. 
{\it Right:} Histogram of the distribution. The green line shows a gaussian fit to this distribution.
}
\label{fig:stat3}
\end{figure*}

As mentioned above, we have 54 cores that are detected in at least
one of the three deuterated species. For those detected in more than
one line we define the core velocity, $v_c$, as an average of the detected centroid velocities, weighted by their measurement uncertainties.


The core velocities $v_c$ are illustrated in \autoref{fig:core_vel}.
For cores with only one \ceighteeno{}(2-1) component, we compare the
difference in centroid velocity between the \ceighteeno{} and $v_c$
directly. If multiple CO velocities are found along line of sight, we
assume the component closest to $v_c$ is the one associated with the
core, following the discussion in \citet{Kirk07}.  This comparison is
shown in \autoref{fig:stat3}.


The median value of the velocity offset is 0.01 \kms, with a standard
deviation of about 0.3~\kms.  The majority of cores (71\%) have core
and envelope velocity offsets less than the sound speed of the ambient
medium (0.27~\kms~ for 20~K temperature).  This percentage is higher
than that in NGC 1333, for which \citet{Walsh07} found half of their
cores have differences greater than the sound speed, but similar to
that seen in the Perseus cloud \citep{Kirk07}. As discussed in
\citet{Walsh04}, small relative motions between cores and envelopes
could be interpreted as an indication of quiescence on small scales
and this would appear to argue against a competitive accretion
scenario for star formation \citep{Bonnell06}, in which dense cores
gain most of their mass by sweeping up material as they move through
the cloud.


\subsection{Virial state of dense cores}

We now examine the dynamical state of the dense cores, i.e., the 
comparison of their internal kinetic energy ($E_K$) and gravitational energy ($E_G$). This ratio is captured by the virial parameter \citep{Bertoldi92}, defined as 
\begin{equation}
 \alpha \equiv 5\sigma_c^2R_c/(GM_{c}) = 2aE_K/|E_G|,
\end{equation}
where $\sigma_c$ is the intrinsic 1D velocity dispersion of the 
core and $R_c$ is the core radius. The dimensionless parameter $a$ accounts for modifications that apply in the case of non-homogeneous and non-spherical density distributions. For a spherical core with a radial density profile that is a power law $\rho\propto r^{-k_\rho}$, then for $k_\rho=0,1,1.5,2$, $a=1,10/9,5/4,5/3$. We adopt a fiducial value of $k_\rho=1.5$ and $a=5/4$, following \citet[][]{Mckee03}.
For a self-gravitating, unmagnetized core without rotation, a virial
parameter above a critical value $\alpha_{\rm cr}= 2a$ indicates that 
the core is unbound and may expand, while one below $\alpha_{cr}$ 
suggests that the core is bound and may collapse. 

We measure core 1D velocity dispersions, $\sigma_c$, from each of the three 
dense gas tracers, i.e., \ntwodp{}, \dcop{} and \dcn. As shown above, their
line widths can vary for the same core, so we calculate the 
virial parameters separately using each tracer.
We derive the intrinsic velocity dispersion from the observed dispersion
following equation (3) (replacing \ceighteeno{} with other species). For the core masses, we use the values estimated assuming a temperature of 20~K, as listed in 
\autoref{table:cores}.



For core radius we attempt two methods. The first is to use the
effective radius calculated from the Dendrogram-returned area (see \S5.1). For the second method we adopt a deconvolved size defined as
$R_{c}$=$\sqrt{(A-A_{\rm beam})/\pi}$, where $A$ and $A_{\rm beam}$
are the core area and synthesised beam size, respectively.  Note that
in our core identification process, we have allowed for cores with an
area smaller than the synthesized beam size.  Here for the virial
analysis we ignore the cores with areas smaller than 1.5$\times A_{\rm
  beam}$, for which the deconvolved sizes could have very large
uncertainties. This criterion excludes 34 out of 76 cores.


\begin{figure*}[ht!]
\epsscale{1.}\plotone{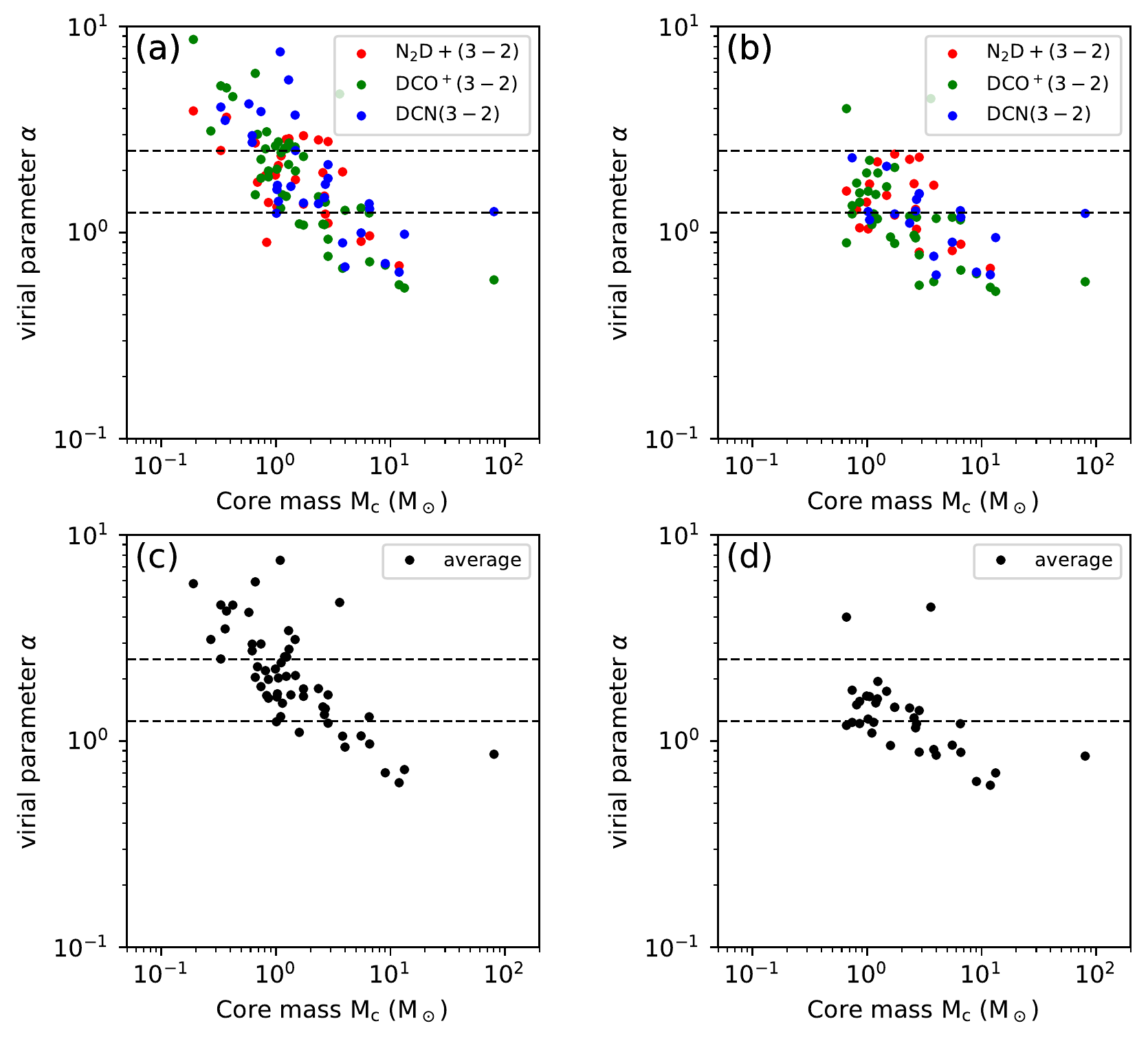}
\caption{
\textit{(a)} Virial parameter versus core mass, with radius measured
from the dendrogram defined area and velocity dispersion measured with
different dense gas tracers, as shown in the legend. The critical
value of $\alpha_{\rm cr}=2a\rightarrow 2.5$ is shown by the upper
dashed line: cores below this line are gravitationally bound. The
lower dashed line shows the virial equilibrium case of
$\alpha=a\rightarrow5/4$.
\textit{(b)} As (a), but with core radius estimated after allowing for
beam deconvolution. Small cores, i.e., with areas $>1.5A_{\rm beam}$
are excluded. \textit{(c)} Same as (a) but we take the linear
average of the non-thermal line width measured via different tracers
to derive an average virial parameter. \textit{(d)} Same as
(c) but using the deconvolved size.}
\label{fig:stat4}
\end{figure*}

\begin{figure*}[ht!]
\epsscale{1.}\plotone{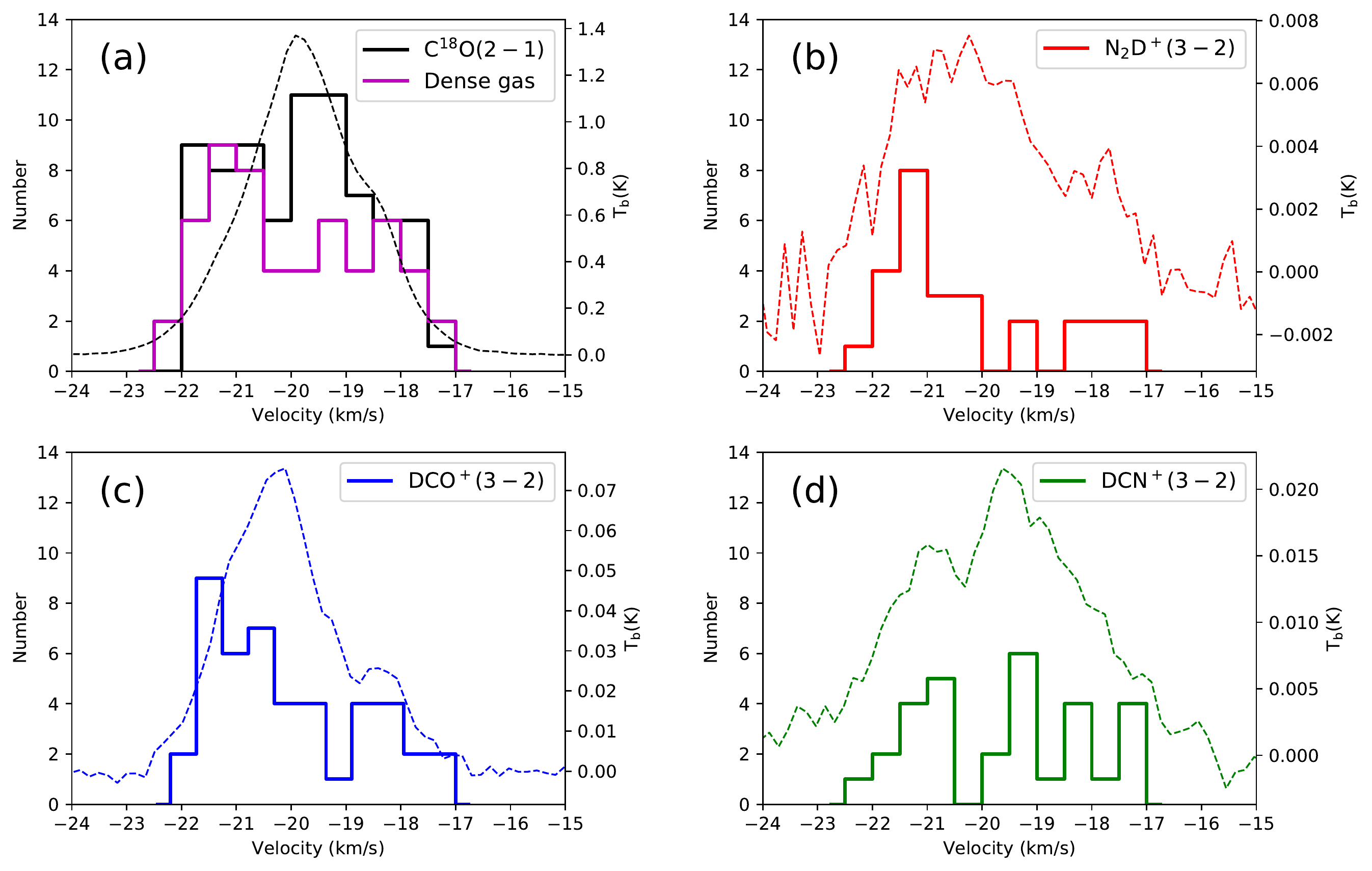}
\caption{
{\it (a)} Distribution of \ceighteeno{}(2-1) core centroid 
velocities (black). Overlaid is the TP \ceighteeno{}(2-1) spectrum (averaged over
2.5\arcmin~ radius aperture) for comparison. The results of core velocities
measured with deuterated species are shown in magenta histogram.
{\it (b)} Same as {\it (a)} but only for \ntwodp{}(3-2).
{\it (c)} Same as {\it (a)} but only for \dcop{}(3-2).
{\it (d)} Same as {\it (a)} but only for \dcn{}(3-2). The TP \dcn{}(3-2) spectrum is averaged
over 1\arcmin~ radius aperture.
}
\label{fig:stat5}
\end{figure*}

\autoref{fig:stat4}a and b display the virial parameters measured with
different tracers versus core mass for the two methods described
above.  In \autoref{fig:stat4}c and d we combine the measurements from
different tracers by taking the linear average of their non-thermal
velocity dispersion in the virial parameter derivation.

We see virial parameters range from 0.5 to 10 as measured by
individual dense gas tracers. There is a trend for more massive
cores to have smaller virial parameters, but this is generally
expected since $\alpha\propto M_c^{-1}$.
The scatter is significantly
reduced for the deconvolved size method, with most measurements
ranging from 0.5 to 3. This suggests most data points with virial
parameter $>5$ in panel (a) could arise from overestimation in the
core radius. We do not find significant systematic differences between
different tracers. The median values are 1.35, 1.19, 1.23 for \ntwodp,
\dcop{} and \dcn, respectively.

The virial parameters estimated by averaging all the available dense
gas data for each core show a further reduction in the scatter. For
the second method with deconvolved sizes that focus on the larger
cores, we obtain a median value of 1.22, and a standard deviation of
0.88. Most cores have a virial 
parameter that is consistent with a value expected in virial equilibrium, 
given the uncertainties.

The uncertainties in the derived virial parameters come from
uncertainties in measured 1D line dispersion $\sigma_{\rm obs}$, mass
and temperature.  The fitting error of $\sigma_{\rm obs}$ is typically
$\lesssim$20\%, resulting in $\sim$ 40\% uncertainty in $\sigma_{\rm
  obs}^2$.  The assumed temperature will systematically affect
estimation of the dense core mass and also the thermal line width
component in equation (3).  For example, with a typical $\sigma_{\rm
  DCO^+}$ = 0.36~\kms, if temperatures of 15~K or 30~K were to be
adopted, then the virial parameters would differ by factors of 0.6 and
1.9, respectively.
Also considering other uncertainties in the mass
estimate, like dust opacity, gas-to-dust mass ratio, dust emission
fluxes, and distances, overall, we estimate the absolute virial
parameter uncertainties to be about a factor of 2.5. However, this uncertainty factor is itself quite uncertain and includes systematic effects, some of which are not expected to vary that much from core to core.

 Given the best case derived average value of $\alpha_{\rm vir}=1.22$, we conclude that the dense cores in G286, which span a wide range of masses from $\lesssim1\:M_\odot$ to $\sim100\:M_\odot$, have properties that are
consistent with them being close to virial equilibrium. In this case, self-gravity
would have been important in forming the cores and a basic
assumption of the turbulent core accretion model of star formation \citep{Mckee03} would be confirmed. However, given the potentially large
systematic uncertainties it is difficult to be more certain about whether the
dense cores are actually closer to a supervirial or subvirial state, or whether
magnetic fields are playing a role in supporting the cores. The situation could be improved in the future with accurate core-scale temperature and magnetic field measurements.



\subsection{Core to core velocity dispersion}

\begin{deluxetable}{lccc}
\tabletypesize{\scriptsize}
\caption{Comparison of core-to-core velocity dispersion and velocity dispersion required for virial equilibrium (in \kms) }
\label{table:comp}
\tablehead{
\colhead{} & \colhead{$\rm \sigma_{c-c,C^{18}O}$} & \colhead{$\rm \sigma_{c-c,deu}$\tablenotemark{a}} & \colhead{$\rm \sigma_{vir}$} 
}
\startdata
blue group  & 0.83$\pm$0.10 & 0.78$\pm$0.10 &  1.48$\pm$0.37   \\
red group  & 0.75$\pm$0.10  & 0.79$\pm$0.12 &  0.96$\pm$0.24   \\
total     & 1.27$\pm$0.11  & 1.39$\pm$0.13  & 1.42$\pm$0.36    \\
\enddata
\tablenotetext{a}{For core velocity measurements combining deuterated species of \ntwodp, \dcop~ and \dcn.}
\end{deluxetable}

The relative motion between dense cores can be quantified using
the core-to-core velocity dispersion $\sigma_{\rm c-c}$, 
i.e., the standard deviation of the core centroid velocities.  It can
be compared with the velocity dispersion of the large scale diffuse
gas out of which these dense cores presumably formed or the initial
velocity dispersion of newborn stars, and as such, provides important
constraints on theoretical models and simulations of star cluster
formation \citep[e.g.,][]{Kirk10,Foster15}.


Here our target G286 offers an interesting case of a massive
protocluster that is still in the gas-dominated phase and actively
forming stars. To measure the core velocity despersion, we show the
core velocity distributions measured with \ceighteeno{}(2-1),
\ntwodp{}(3-2), \dcop{}(3-2) and \dcn{}(3-2) in
\autoref{fig:stat5}. For comparision the large scale total power
spectra of each line are also overlaid. The results combining
velocities measured with deuterated tracers (54 cores, see
\autoref{sec:para}) are also displayed in \autoref{fig:stat5}(a).  We
then calculate the standard deviation of these distributions,
obtaining 1.27 $\pm$ 0.11 \kms~ for the \ceighteeno-detected sample,
1.52 $\pm$ 0.21 \kms~ for the \ntwodp{} sample, 1.40 $\pm$ 0.15 \kms~
for the \dcop{} cores, 1.50 $\pm$ 0.20 \kms~ for \dcn{} cores and 1.39
$\pm$ 0.13 \kms~ for the combined results. The uncertainties here only
account for sampling errors due to limited sample size, assuming the
data points are drawn from a normal distribution.

In contrast to previous results in nearby cluster-forming clouds like
Ophiuchus and Perseus \citep[][]{Andre07,Kirk07,Kirk10}, our core
velocities cover a wide range from -22.5 to -17 \kms~ and the
distribution is not well approximated with a single gaussian
component. This is particularly clear for \ntwodp{} and \dcop{}: for
these two tracers the core velocities exhibit a bimodal distribution
with two velocity groups, which agrees well with the averaged TP
spectra. \dcn{} picks up core velocites in a relatively uniform
pattern, filling in the gap around -19 \kms~, and hence the
distribution combining all the deuterated species is more flat, though
more cores still cluster in the ``blue'' group at $\sim$ -21 \kms. On
the other hand, though the \ceighteeno{} profile can be characterized
with a gaussian (with some skewness) peaking aroud -20 \kms~ and we do
have more \ceighteeno-detected cores close to the systemic velocity
(-20 \kms~ to -19 \kms), the \ceighteeno-measured core velocity
distribution is still relatively flat.  This indicates that the core
to core velocity dispersion we measured here is largely contributed by
the global velocity patterns.

The core velocity dispersion $\sigma_{\rm c-c}$ can be compared with
the dispersion required for virial equilibrium on the protocluster
clump scale $\sigma_{\rm cl,vir}$, and its actual gas velocity
dispersion, $\sigma_{\rm cl}$.
For $\sigma_{\rm cl,vir}$ we again follow \citet{Bertoldi92}:
\begin{equation}
   \sigma_{\rm cl,vir} = \sqrt {\frac{a G M_{\rm cl}}{5R_{\rm cl}}}. 
\end{equation}
As with cores, we again adopt $k_\rho=1.5$, so that $a=5/4$. We
choose a size of $R_{\rm cl} = 1.54$~pc, which is the geometric
mean of the major and minor axes of the large ellipse boundary shown in \autoref{fig:overview}.
From the {\it Herschel}-SED-derived mass surface density map 
we obtain a mass for this region of
$\sim 2900\:M_\odot$. Thus, $\sigma_{\rm cl,vir}=1.42 \pm 0.36$~\kms, where
the error comes assuming a 50\% uncertainty in the mass estimate. 
We also tried a smaller aperture (by using major and minor axes that are a factor
of 1/$\sqrt{2}$ smaller than the current ones) to more closely encompass the region containing dense cores,
which gives $\sigma_{\rm cl,vir}=1.54 \pm 0.39$~\kms.
The
mass here only accounts for the gas component, since we do not expect
significant contribution from stellar mass: \citet{Andersen17}
estimated a total current stellar mass of $\sim 240 M_\odot$ in a
similarly sized region.  Thus the observed values of $\sigma_{\rm
  c-c}$ are comparable or slightly smaller than $\sigma_{\rm cl,vir}$, depending on which
  tracer is used.

For $\sigma_{\rm cl}$, we measure the line width of average TP spectra
of \ceighteeno{}(2-1) in this region.  The purpose here is to compare
core-to-core motions with the spread of motions seen over the region
as a whole to reveal how connected the dense cores are to the larger
scale gas in the region.
A gaussian fit of the \ceighteeno{}(2-1) line gives $ \sigma_{\rm
  cl,C^{18}O}$ = 1.09 $\pm$ 0.01 \kms. To account for the thermal
component we correct this value following equation (3) assuming a
temperature of 20~K and obtain $ \sigma_{\rm cl}$ = 1.12 \kms.  

In summary, the 1D dispersion measured in gas tracers, $\sigma_{\rm
  cl}$ (1.12 $\pm$ 0.01 \kms), is slightly smaller, but still 
  consistent with $\sigma_{\rm cl,vir}$ (1.48
$\pm$ 0.37 \kms), indicating that the G286 clump could be in approximate virial equilibrium (assuming it is a single, coherent dynamical system) or modestly sub-virial, but it is hard to be more precise given the uncertainties. We also see
a range of $\sigma_{\rm c-c}$ values using different tracers and 
the value using \ceighteeno(2-1), which includes most cores in this sample, i.e., 
$\sigma_{\rm c-c,C^{18}O}$ = 1.27 $\pm$ 0.11 \kms, is similar to $\sigma_{\rm cl,vir}$,
but slight larger than $\sigma_{\rm cl}$.
Here the dense core velocity distribution is more flat, while both the
\ceighteeno{} core velocities and the TP \ceighteeno{} spectrum cover
a similar velocity range. This means there is a deficiency of dense cores near
the systemic velocity ($\sim$ 20 \kms), where the bulk of the
\ceighteeno-traced gas is located. This deficiency is clearer in the
distributions traced by \ntwodp{} and \dcop{}, and hence an even
larger $\sigma_{\rm c-c}$ is measured with these two tracers. The two
velocity groups seen in \ntwodp{} and \dcop{} (at $\sim$ -21 \kms and
-18\kms) are actually spatially distinct (see \autoref{fig:mom01},
\autoref{fig:12m_map},\autoref{fig:core_vel}), with the more redshifted
cores mostly located in the NE-SW filament and the more blueshifted cores
in the NW-SE filament and the E-W filament. A similar velocity pattern
is also seen for \ceighteeno{} in \autoref{fig:mom01}, indicating 
the dense cores are still well coupled
with the large-scale motions within the cloud.

To better characterize the core-to-core velocity dispersions of the
two velocity groups, we adopt a simple boundary to spatially differentiate 
them, shown as the dashed green line in \autoref{fig:core_vel}b. The blueshifted
subsample (northwest of the boundary) has a core-to-core velocity
disperion $\sigma_{\rm c-c,blue}$ of 0.83$\pm$0.10 \kms~ for \ceighteeno-detected
cores (or 0.78$\pm$0.10\kms~ for measurements combining \ntwodp, \dcop~ and \dcn).
Similarly, we have $\sigma_{\rm c-c,red}$ = 0.75$\pm$0.10 \kms~ for 
\ceighteeno~ and 0.79$\pm$0.12 \kms~ for combined dense gas results. For the dispersion 
required for virial equilibrium, $\sigma_{\rm cl,vir}$, we estimate the mass
and radius using two approximate ellipse boundaries for each velocity group (shown in \autoref{fig:overview}),
 which yields $\sigma_{\rm cl,vir,red}$ = 0.96$\pm$0.24 \kms, and $\sigma_{\rm cl,vir,blue}$ = 1.48$\pm$0.37 \kms. These numbers are also summarized in \autoref{table:comp}.
Therefore, at least for the ``blue'' group, the core velocity dispersion 
$\sigma_{\rm c-c}$ is appears to be smaller than $\sigma_{\rm cl,vir}$, potentially indicating it is kinematically cold and sub-virial, perhaps due to coherent motions within a filament. We also see that the
dispersion in core velocities for the whole cloud mainly arises from the
global velocity pattern of the red and blue groups.

The origin of this velocity pattern is uncertain. In the filament
collapse scenario, as observed in some hub-filament systems, accretion
flows are channeling gas to the junctions where star formation is
often most active \citep[e.g.,][]{Kirk13,Peretto14,Liu16}.  It is
possible that these converging flows are reflected in different LOS
velocities depending on the 3D configurations. We will presumably have
more massive cores in the hub region (near the systemic velocity), but
not necessarily a larger number of cores, as suggested by our
observations. Smoothly varying velocities along filaments is expected
in this picture. We do see indications of a velocity gradient of dense
cores along the filaments, but it is not clear in \ceighteeno{}, for
which the spectra are often complex. Further higher sensitivity
observations of \ntwohp{} and \ammonia{} will help investigate the gas
velocity gradient along filaments.


Alternatively, the two main velocity components seen in \ntwodp{} and
\dcop{} could be tracing two interacting clouds/filaments, with the
central region as the collision interface
\citep[e.g.,][]{Nakamura14}. Such a mechanism could be consistent with
a larger-scale cloud-cloud collision scenario that has been reported
in other star-forming regions
\citep[e.g.,][]{Furukawa09,Fukui14,Gong17}.

\citet{Andersen17} analysed the stellar population in G286 and found
evidence for at least three different sub-clusters associated with the
molecular clump based on differences in extinction and disk
fractions. It is unclear how the dense gas distribution and ongoing
cluster formation might be related with these past star formation
events.  Future studies of the radial velocity of optically revealed
stars, e.g., the velocity dispersion and its distribution will be of
great interet to understand the cluster formation in G286.
 
\section{Conclusion}\label{sec:conclusion}

We have studied the gas kinematics and dynamics of the massive protocluster
G286.21+0.17 with ALMA using spectral lines of \ceighteeno(2-1),
\ntwodp(3-2), \dcop(3-2) and DCN(3-2). The main results are as
follows:

\begin{enumerate}

\item Morphologically, \ceighteeno{}(2-1) traces more extended
  emission, while \ntwodp{}(3-2) and \dcop{}(3-2) are more closely
  associated with the dust continuum. DCN(3-2) is strongly
  concentrated towards the protocluster center, where no or only weak
  detection is seen for \ntwodp{} and \dcop, possibly due to a
  relatively evolved evolutionary stage in the central region
  involving chemical evolution at higher temperatures.

\item Based on 1.3~mm continuum, G286 is composed of several pc-scale
  filamentary structures: the NE-SW filament in northwest, and the
  NW-SE filament in the southeast, as well as another filament
  oriented in the E-W direction that is more clearly seen in \dcop.
  The NE-SW filament is associated with redshifted \ceighteeno{}
  emission while the NW-SE and E-W filament are mainly associated with
  blueshifted gas. Other tracers show similar velocity structures.

\item We performed a filamentary virial analysis towards the NE-SW
  filament.  We divided the filament into four strips and the values
  of $m_f/m_{\rm vir,f}$ of the four strips range from 0.60 to
  2.39. Within the uncertainties, these values are consistent with the
  filament being in virial equilibrium, without accounting for surface
  pressure and magnetic support terms.  We also detected a steady
  velocity gradient of 2.84 $\rm km^{-1} pc^{-1}$ along the
  filament, which may arise from infall motion.

\item We analysed the spectra of 74 continuum dense cores and
  measureed their centroid velocities and internal velocity
  dispersions.  There are no statistically significant velocity
  offsets among different tracers. \ceighteeno{} has systematically
  larger velocity dispersion compared with other tracers.

\item The majority (71\%) of the dense cores have subthermal velocity
  offsets with respect to their surrounding \ceighteeno{} emitting
  envelope gas, similar as found in previous studies for low-mass star
  formation environments \citep[e.g.,][]{Kirk07}.

\item  We measured the virial parameters of the dense core
  in G286, which span more than two orders of magnitude in mass. The average value of these virial parameters is close to unity, suggesting the cores are close
  to virial equilibrium and self-gravity has been important for forming the cores. However, this conclusion is subject to revision if there is a large systematic error in the mass estimates of the cores.

\item The core to core velocity dispersion in G286 is similar to that
  required for virial equilibrium in the protocluster potential, but
  with the velocity distribution largely composed of two spatially
  distinct velocity groups. This indicates that the dense molecular
  gas has not yet relaxed to virial equilibrium in the protocluster potential, even though the total velocity dispersion would indicate such a condition. The same analysis for the sub-regions
  corresponding to each velocity group reveals smaller core to core velocity 
  dispersions, with one case consistent with virial equilibrium and the other case indicating a sub-virial state.

\end{enumerate}

\acknowledgements This paper makes use of the following ALMA data:
ADS/JAO.ALMA\#2015.1.00357.S. ALMA is a partnership of ESO
(representing its member states), NSF (USA) and NINS (Japan), together
with NRC (Canada), NSC and ASIAA (Taiwan), and KASI (Republic of
Korea), in cooperation with the Republic of Chile. The Joint ALMA
Observatory is operated by ESO, AUI/NRAO, and NAOJ. The National Radio
Astronomy Observatory is a facility of the National Science Foundation
operated under cooperative agreement by Associated Universities, Inc.

\bibliographystyle{aasjournal}                                                   
\bibliography{refer}

\acknowledgments

\newpage

\appendix 

\section{Properites of dense cores in G286}

\startlongtable
\begin{deluxetable*}{ccccccccc}
\tabletypesize{\scriptsize}
\renewcommand{\arraystretch}{0.9}
\tablecaption{Estimated physical parameters for 1.3 mm continuum cores}
\tablehead{
\colhead{core} & \colhead{Ra} & \colhead{Dec} & \colhead{$I_{peak}$} & \colhead{$S_\nu$} & \colhead{$M_c$}  & \colhead{$R_c$} & \colhead{$\Sigma_c $} & \colhead{$n_{H,c}$ } \\
\colhead{} & \colhead{($^{\circ}$)} & \colhead{($^{\circ}$)} & \colhead{mJy beam$^{-1}$} & \colhead{(mJy)} & \colhead{$M_\odot$}  & \colhead{(0.01pc)} & \colhead{(g cm$^{-2}$)} & \colhead{10$^6$cm$^{-3}$ }
}
\startdata
1  &  159.63383  &  -58.31897  &  60.14  &  420.29  &  80.24  &  3.63  &  4.068  &  11.71  \\ 2  &  159.63524  &  -58.32063  &  15.32  &  47.19  &  9.01  &  1.74  &  1.994  &  11.99  \\
3  &  159.64045  &  -58.32043  &  11.20  &  34.13  &  6.52  &  1.92  &  1.179  &  6.41  \\
4  &  159.64373  &  -58.32201  &  11.10  &  34.49  &  6.58  &  1.76  &  1.416  &  8.39  \\
5  &  159.63973  &  -58.32167  &  11.06  &  69.42  &  13.25  &  2.71  &  1.209  &  4.66  \\
6  &  159.63154  &  -58.31674  &  10.03  &  14.94  &  2.85  &  1.05  &  1.713  &  16.97  \\
7  &  159.64328  &  -58.32094  &  9.43  &  29.04  &  5.54  &  1.68  &  1.308  &  8.12  \\
8  &  159.63153  &  -58.31933  &  9.04  &  7.07  &  1.35  &  0.73  &  1.694  &  24.24  \\
9  &  159.64708  &  -58.32530  &  8.73  &  20.00  &  3.82  &  1.43  &  1.244  &  9.07  \\
10  &  159.63546  &  -58.32133  &  8.52  &  5.40  &  1.03  &  0.66  &  1.569  &  24.74  \\
11  &  159.63163  &  -58.31720  &  8.38  &  3.23  &  0.62  &  0.51  &  1.574  &  32.16  \\
12  &  159.63177  &  -58.31842  &  8.30  &  5.69  &  1.09  &  0.68  &  1.570  &  24.11  \\
13  &  159.63045  &  -58.31534  &  8.29  &  14.93  &  2.85  &  1.34  &  1.061  &  8.28  \\
14  &  159.63572  &  -58.31830  &  7.92  &  5.31  &  1.01  &  0.70  &  1.380  &  20.58  \\
15  &  159.66617  &  -58.32238  &  7.90  &  13.47  &  2.57  &  1.56  &  0.710  &  4.77  \\
16  &  159.63329  &  -58.32012  &  7.57  &  7.69  &  1.47  &  0.83  &  1.429  &  18.01  \\
17  &  159.62948  &  -58.31815  &  7.32  &  6.79  &  1.30  &  0.81  &  1.325  &  17.11  \\
18  &  159.66145  &  -58.32416  &  7.31  &  9.14  &  1.74  &  1.26  &  0.740  &  6.16  \\
19  &  159.63511  &  -58.31409  &  6.87  &  62.38  &  11.91  &  3.09  &  0.833  &  2.82  \\
20  &  159.63008  &  -58.31798  &  6.67  &  6.77  &  1.29  &  0.83  &  1.258  &  15.86  \\
21  &  159.63146  &  -58.31591  &  6.52  &  12.32  &  2.35  &  1.23  &  1.045  &  8.90  \\
22  &  159.64112  &  -58.31903  &  6.45  &  21.01  &  4.01  &  1.79  &  0.835  &  4.87  \\
23  &  159.62922  &  -58.31562  &  6.31  &  14.14  &  2.70  &  1.36  &  0.975  &  7.49  \\
24  &  159.63281  &  -58.31702  &  6.16  &  3.02  &  0.58  &  0.58  &  1.144  &  20.61  \\
25  &  159.64503  &  -58.32397  &  6.12  &  13.86  &  2.65  &  1.44  &  0.856  &  6.23  \\
26  &  159.63331  &  -58.31758  &  6.08  &  3.27  &  0.62  &  0.60  &  1.160  &  20.21  \\
27  &  159.63752  &  -58.31851  &  6.06  &  8.92  &  1.70  &  1.13  &  0.896  &  8.30  \\
28  &  159.64744  &  -58.32461  &  5.79  &  5.82  &  1.11  &  0.87  &  0.987  &  11.89  \\
29  &  159.64422  &  -58.32289  &  5.28  &  4.33  &  0.83  &  0.79  &  0.875  &  11.52  \\
30  &  159.64468  &  -58.32329  &  5.22  &  3.61  &  0.69  &  0.73  &  0.866  &  12.39  \\
31  &  159.62961  &  -58.31916  &  5.13  &  6.24  &  1.19  &  0.91  &  0.968  &  11.16  \\
32  &  159.63175  &  -58.32042  &  5.12  &  5.04  &  0.96  &  0.89  &  0.814  &  9.57  \\
33  &  159.62987  &  -58.32137  &  4.50  &  12.73  &  2.43  &  1.45  &  0.769  &  5.53  \\
34  &  159.64877  &  -58.32960  &  4.43  &  7.77  &  1.48  &  1.34  &  0.555  &  4.34  \\
35  &  159.63103  &  -58.32106  &  4.16  &  2.89  &  0.55  &  0.71  &  0.724  &  10.60  \\
36  &  159.62937  &  -58.32062  &  4.12  &  2.70  &  0.52  &  0.68  &  0.745  &  11.45  \\
37  &  159.63976  &  -58.32276  &  4.02  &  8.54  &  1.63  &  1.45  &  0.516  &  3.71  \\
38  &  159.63000  &  -58.32016  &  4.00  &  5.38  &  1.03  &  0.96  &  0.737  &  7.99  \\
39  &  159.63706  &  -58.31779  &  3.92  &  3.64  &  0.70  &  0.87  &  0.614  &  7.38  \\
40  &  159.62838  &  -58.32134  &  3.86  &  2.22  &  0.42  &  0.62  &  0.731  &  12.26  \\
41  &  159.63368  &  -58.31362  &  3.81  &  5.43  &  1.04  &  1.07  &  0.610  &  5.98  \\
42  &  159.62900  &  -58.31655  &  3.77  &  6.43  &  1.23  &  1.16  &  0.612  &  5.52  \\
43  &  159.64888  &  -58.32789  &  3.77  &  5.20  &  0.99  &  1.08  &  0.568  &  5.49  \\
44  &  159.64251  &  -58.31957  &  3.69  &  6.50  &  1.24  &  1.12  &  0.660  &  6.15  \\
45  &  159.63874  &  -58.31673  &  3.64  &  8.38  &  1.60  &  1.44  &  0.514  &  3.72  \\
46  &  159.67328  &  -58.32603  &  3.63  &  5.34  &  1.02  &  1.17  &  0.501  &  4.49  \\
47  &  159.63370  &  -58.31682  &  3.61  &  1.91  &  0.36  &  0.62  &  0.643  &  10.93  \\
48  &  159.64524  &  -58.32276  &  3.46  &  3.46  &  0.66  &  0.90  &  0.549  &  6.40  \\
49  &  159.64849  &  -58.32509  &  3.43  &  3.86  &  0.74  &  0.91  &  0.599  &  6.90  \\
50  &  159.63991  &  -58.32611  &  3.42  &  4.53  &  0.86  &  1.11  &  0.472  &  4.45  \\
51  &  159.67270  &  -58.32482  &  3.38  &  9.11  &  1.74  &  1.56  &  0.477  &  3.20  \\
52  &  159.64614  &  -58.32435  &  3.32  &  1.72  &  0.33  &  0.60  &  0.608  &  10.60  \\
53  &  159.64167  &  -58.31675  &  3.28  &  18.88  &  3.60  &  2.32  &  0.449  &  2.03  \\
54  &  159.64989  &  -58.32882  &  3.27  &  4.23  &  0.81  &  1.00  &  0.545  &  5.72  \\
55  &  159.64225  &  -58.32300  &  3.23  &  1.51  &  0.29  &  0.58  &  0.571  &  10.29  \\
56  &  159.63973  &  -58.32419  &  3.18  &  3.90  &  0.74  &  0.98  &  0.517  &  5.50  \\
57  &  159.63632  &  -58.32468  &  3.12  &  3.64  &  0.69  &  1.01  &  0.460  &  4.78  \\
58  &  159.65086  &  -58.32812  &  2.98  &  3.46  &  0.66  &  0.99  &  0.454  &  4.82  \\
59  &  159.63897  &  -58.32474  &  2.94  &  5.99  &  1.14  &  1.23  &  0.502  &  4.25  \\
60  &  159.63765  &  -58.32215  &  2.90  &  2.61  &  0.50  &  0.88  &  0.432  &  5.14  \\
61  &  159.63372  &  -58.31557  &  2.87  &  1.43  &  0.27  &  0.63  &  0.460  &  7.63  \\
62  &  159.63206  &  -58.32324  &  2.84  &  1.69  &  0.32  &  0.68  &  0.467  &  7.17  \\
63  &  159.64094  &  -58.30845  &  2.83  &  5.51  &  1.05  &  1.25  &  0.451  &  3.77  \\
64  &  159.67355  &  -58.32445  &  2.70  &  1.74  &  0.33  &  0.72  &  0.425  &  6.13  \\
65  &  159.64838  &  -58.31984  &  2.68  &  5.75  &  1.10  &  1.31  &  0.425  &  3.38  \\
66  &  159.64144  &  -58.32606  &  2.67  &  4.48  &  0.86  &  1.17  &  0.419  &  3.75  \\
67  &  159.63598  &  -58.31504  &  2.64  &  1.38  &  0.26  &  0.64  &  0.429  &  6.98  \\
68  &  159.67361  &  -58.32257  &  2.61  &  1.94  &  0.37  &  0.76  &  0.431  &  5.93  \\
69  &  159.64169  &  -58.32479  &  2.60  &  1.86  &  0.36  &  0.74  &  0.430  &  6.05  \\
70  &  159.63645  &  -58.32321  &  2.56  &  6.47  &  1.23  &  1.39  &  0.424  &  3.18  \\
71  &  159.63106  &  -58.32798  &  2.56  &  1.49  &  0.29  &  0.66  &  0.434  &  6.84  \\
72  &  159.63961  &  -58.30961  &  2.55  &  1.86  &  0.36  &  0.75  &  0.427  &  5.99  \\
73  &  159.63557  &  -58.31262  &  2.52  &  1.20  &  0.23  &  0.58  &  0.460  &  8.35  \\
74  &  159.63958  &  -58.31643  &  2.49  &  1.31  &  0.25  &  0.62  &  0.436  &  7.35  \\
75  &  159.64786  &  -58.32908  &  2.47  &  0.99  &  0.19  &  0.53  &  0.444  &  8.69  \\
76  &  159.63002  &  -58.32627  &  2.39  &  1.16  &  0.22  &  0.59  &  0.423  &  7.46  \\
\enddata
\label{table:cores}
\end{deluxetable*}

\section{Spectra fitting of the core sample}

\begin{figure*}[ht!]
\epsscale{1.}\plotone{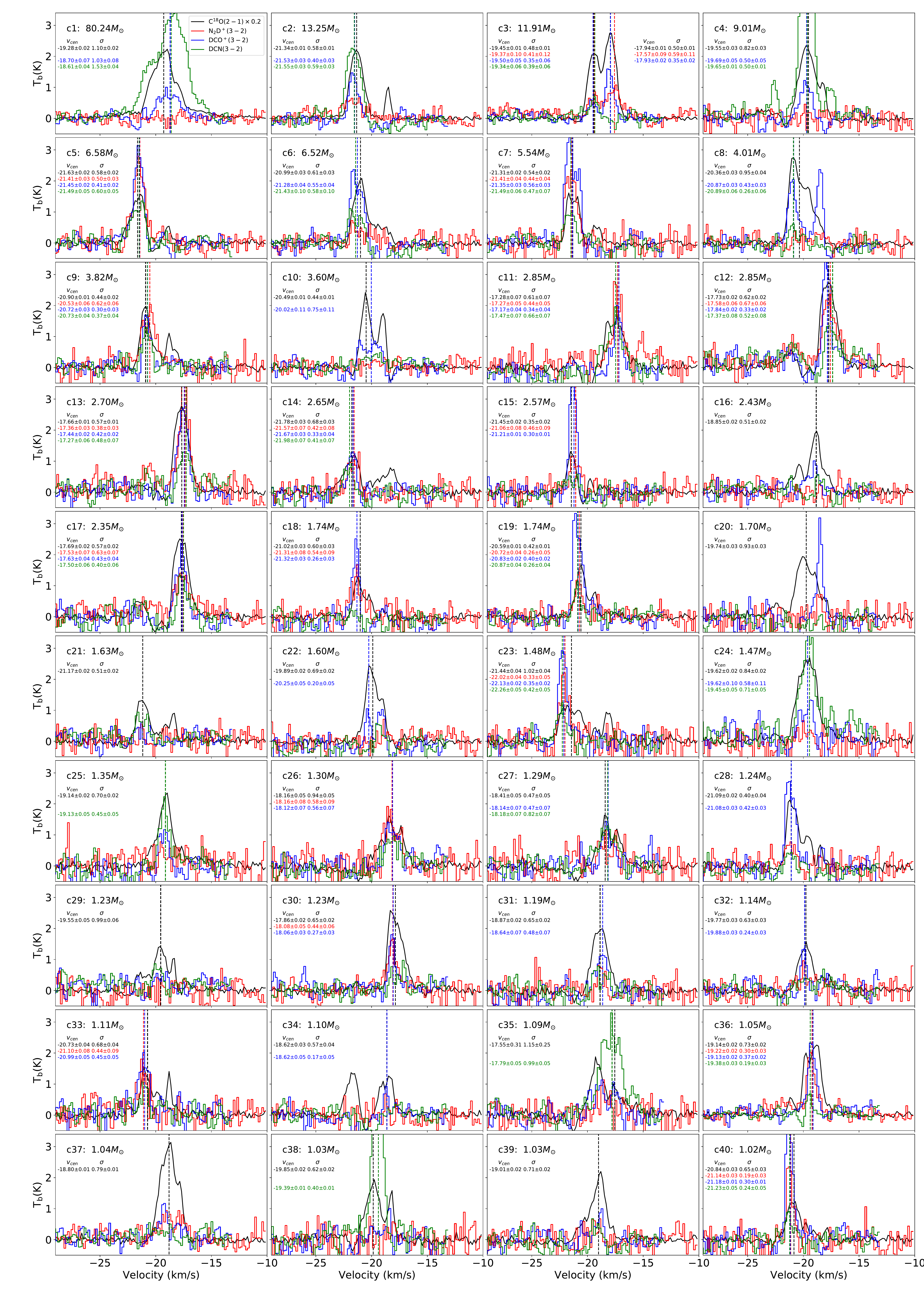}
\caption{
Spectra of \ceighteeno{}(2-1), \ntwodp{}(3-2), \dcop{}(3-2) and \dcn{}(3-2) of 76 continuum cores shown in different colors. The core masses are labeled on the top left. For the spectrum with a 
peak flux greater than 4 $\sigma$ we perform a gaussian fitting.
The returned parameters (centroid velocity,
velocity dispersion) for each line are displayed on the top left, in 
the same color as the corresponding line.
The dashed vertical lines indicate the centroid velocity from line fitting
(If there are multiple components for \ceighteeno, only the main component (the one closer to other dense gas
tracers, see text) is shown).
}
\label{fig:cspec1}
\end{figure*}

\begin{figure*}[ht!]
\epsscale{1.}\plotone{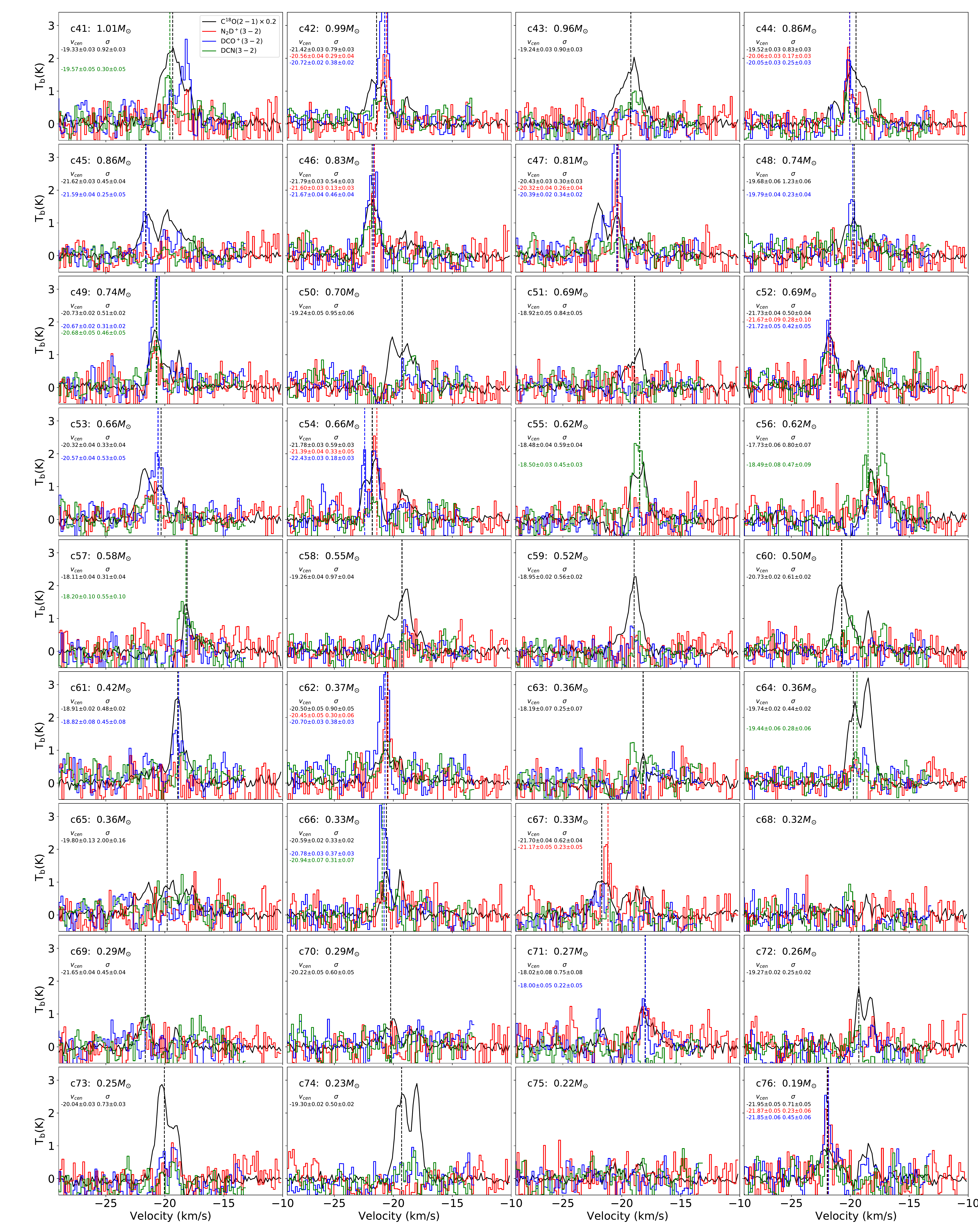}
\caption{
Continue of \autoref{fig:cspec1}.
}
\label{fig:cspec2}
\end{figure*}

\end{document}